\documentclass[12pt]{article} \input epsf.tex
\usepackage{graphicx}
\newcommand{\f}{u}
\renewcommand{\theequation}{\thesection.\arabic{equation}}
\newcommand{\be}{\begin{equation}}
\newcommand{\ee}{\end{equation}}
\newcommand{\bear}{\begin{eqnarray}}
\newcommand{\eear}{\end{eqnarray}} \newcommand{\ba}{\begin{array}}
\newcommand{\ea}{\end{array}}
\newcommand{\lae}{\begin{array}{c}\,\sim\vspace{-21pt}\\<
\end{array}}
\newcommand{\gae}{\begin{array}{c}\,\sim\vspace{-21pt}\\>
\end{array}}
 \newcommand{\CQ}{{\cal Q}}
\newcommand{\CU}{{\cal U}} \newcommand{\CD}{{\cal D}}
\newcommand{\CL}{{\cal L}} \newcommand{\CE}{{\cal E}}
\newcommand{\CN}{{\cal N}} \newcommand{\CH}{{\cal H}}
\begin{document}

\pagestyle{empty} \begin{titlepage}
\def\thepage {} % Kill page numbering

\title{\Large \bf
Neutrinos Vis-\`{a}-vis the Six-Dimensional \\ [.3cm] Standard Model
\\ [1.3cm]}

\author{\normalsize
\bf \hspace*{-.3cm} Thomas Appelquist,
Bogdan A.~Dobrescu, Eduardo Pont\'{o}n, Ho-Ung Yee
 \\ \\ {\small {\it
\vspace*{-5cm}
Department of Physics, Yale University, New
Haven, CT 06520, USA
}}\\
 }

\date{ } \maketitle

\vspace*{-7.4cm}
\noindent \makebox[12.7cm][l]{\small \hspace*{-.2cm}
hep-ph/0201131} {\small YCTP-P1-02 } \\
\makebox[11.8cm][l]{\small \hspace*{-.2cm} January 14, 2002 }
{\small } \\

 \vspace*{10.9cm}

  \begin{abstract}
{\small We examine the origin of neutrino masses and oscillations
in the context of the six-dimensional standard model. The
space-time symmetries of this model explain proton stability and
forbid Majorana neutrino masses. The consistency of the
six-dimensional theory requires three right-handed neutrinos, and
therefore Dirac neutrino masses are allowed. We employ the idea
that the smallness of these masses is due to the propagation of the
right-handed neutrinos in a seventh, warped dimension. We argue
that this class of theories is free of gravitational anomalies.
Although an exponential hierarchy arises between the neutrino
masses and the electroweak scale, we find that the mass hierarchy
among the three neutrino masses is limited by higher-dimension
operators. All current neutrino oscillation data, except for the
LSND result, are naturally accommodated by our model. In the case
of the solar neutrinos, the model leads to the large mixing angle,
MSW solution. The mechanism employed, involving three right-handed
neutrinos coupled to a scalar in an extra dimension, may explain
the features of the neutrino spectrum in a more general class of
theories that forbid Majorana masses.
 }
\end{abstract}

\vfill \end{titlepage}

\baselineskip=18pt \pagestyle{plain} \setcounter{page}{1}

%%%%%%%%%%%%%%%%%%%%%%%%%%%%%%%%%%%%%%%%%%%%%%%%%%%%%%%%%%%%%%%%%%%%
\section{Standard Model in Six Dimensions} \setcounter{equation}{0}

The proposal \cite{Appelquist:2000nn} that all the standard model
fields access extra spatial dimensions above some energy scale
(``universal extra dimensions") has received considerable attention
during the past year.  Precision electroweak measurements require only
that the compactification scale of universal extra dimensions be above
a few hundred GeV, opening up a potentially rich set of signatures,
both in additional precision measurements \cite{Agashe:2001ra, Rizzo:2001sd}
and in
collider searches \cite{Appelquist:2000nn, Rizzo:2001sd, Cheng:2001}.

An especially attractive possibility is that there exist two universal
extra dimensions.  The six-dimensional standard model is chiral, and
the constraints from Lorentz invariance and anomaly cancellation have
remarkable consequences.  The quarks ($\CQ, \CU, \CD$) and leptons
($\CL, \CE$) are four-component Weyl fermions of definite chirality,
labeled by $+$ and $-$.  The cancellation of irreducible gauge
anomalies imposes one of the following two chirality assignments
consistent with Lorentz invariant Yukawa couplings: $\CQ_+, \CU_-,
\CD_-, \CL_\mp, \CE_\pm$, where generational indices are implicit
\cite{Dobrescu:2001ae}. The reducible gauge anomalies can be
canceled via the Green-Schwarz mechanism as discussed in
\cite{Arkani-Hamed:2000hv,Dobrescu:2001ae,Fabbrichesi:2001fx,Borghini:2001sa}.
Gravitational anomaly cancellation requires that each generation
include a gauge singlet fermion $\CN_\pm$ with six-dimensional
chirality opposite to that of the lepton doublet
\cite{Arkani-Hamed:2000hv}.  In addition, the six-dimensional
standard model is the only known theory that constrains the number
of fermion generations to be $n_g
= 3 \; {\rm mod} \; 3$, based on the global anomaly cancellation
condition \cite{Dobrescu:2001ae}.

The two universal extra dimensions have to be compactified on an
orbifold, so that each of the six-dimensional chiral fermions gives in
the effective four-dimensional theory either a left- or a right-handed
zero-mode fermion.  The simplest orbifold compactifications are either
the square $T^2/Z_2$ or $T^2/Z_4$ orbifolds.

An intriguing feature of the six-dimensional standard model is that
the combination of its Lorentz and gauge symmetries can lead to a
sufficient conservation of baryon number, even with the scale of
baryon-number violating physics as low as the TeV range
\cite{Appelquist:2001mj}.  For the $T^2/Z_4$ orbifold, a $Z_8$
subgroup of the six-dimensional Lorentz symmetry is exactly
preserved. In the case of the square $T^{2}/Z_{2}$ orbifold, the
same is true provided the two orbifold fixed points that are
exchanged by a $90^\circ$ rotation in the compactified (transverse)
dimensions are physically indistinguishable. The $Z_8$ symmetry
requires that the baryon and lepton numbers, $\Delta B$ and $\Delta
L$, of any operator in the low-energy four-dimensional Lagrangian
obey the selection rule\footnote{The cancellation of anomalies via
the Green-Schwarz mechanism requires a (four-dimensional) scalar
field that transforms nontrivially under the $Z_8$. Operators that
involve this field can be induced by four-dimensional instanton
effects and could result in a violation of the selection rule
Eq.~(\ref{selectionrule}). We expect these effects to be
negligible. We thank E.~Poppitz for discussions on this point.}
\be
\label{selectionrule}
3\Delta B + \Delta L = 0\; {\rm mod}\, 8 .
\ee
As a result, the proton is very long lived (all $\Delta B = 1$
transitions are governed by very high-dimension operators, and are
therefore strongly suppressed), while neutron--anti-neutron
oscillations ($\Delta B
=2$, $\Delta L =0$) are forbidden.  In the lepton sector, there are
no neutrino Majorana masses\footnote{The fact that the $Z_8$
symmetry forbids Majorana masses was not taken into account in
Ref.~\cite{Fabbrichesi:2001fx}. We note that even if the two
universal extra dimensions were compactified on an arbitrary
$T^2/Z_2$ orbifold, an exact $Z_4$ symmetry would still have
prevented any Majorana mass.}, and more generally neutrino-less
double beta decays ($\Delta B =0$, $\Delta L =2$) are forbidden.
The absence of Majorana masses follows from the properties of the
gamma matrices in six dimensions, namely that the charge
conjugation operator does not flip the chirality.

In this paper we study the implications for neutrino physics of the
six-dimensional standard model.  The mass matrix for the zero-mode
neutrinos is induced dominantly by the following dimension-seven
Yukawa terms in the six-dimensional Lagrangian:
\be - \overline{\CL}_-^i \hat{\lambda}_\CN^{ii^\prime}
\CN_+^{i^\prime}
i\sigma_2 \CH^* +
{\rm h.c.},
\label{lagrangian-6DN}
\ee
where $i,i^\prime $ label the generations, $\CH$ is the six-dimensional
Higgs doublet, and where we have taken the six-dimensional chirality of
$\CL$ to be $-$.  The ensuing Dirac masses of the three neutrino
flavors can accommodate the neutrino oscillation data and all other
experimental constraints, with the exception of the LSND result
\cite{Mills:2001tq}.  It is nevertheless difficult to explain why the
eigenvalues of the Yukawa matrix $\hat{\lambda}_\CN$ are extremely
small.

Since the standard model in extra dimensions is an effective theory,
breaking down at some scale $M_s$ in the TeV range, it is natural to
expect gravity to be strongly coupled there as well.  A structure that
accommodates the observed weakness of the gravitational interaction
should then be added to the universal extra dimensions.  The simplest
possibilities are that either some number of additional flat
dimensions \cite{Arkani-Hamed:1998rs} or one additional warped
dimension \cite{Randall:1999ee} are transverse to the universal ones
and are not accessible to standard model fields.  Each of these
alternatives also provides a possible mechanism for explaining small
but finite Dirac neutrino masses, as first proposed in
\cite{Arkani-Hamed:1998vp} and \cite{Grossman:2000ra}, respectively,
by letting the gauge singlet fermions propagate in these extra
dimension(s).

We concentrate here on the possibility that the singlet fermions,
along with gravity, propagate in a single additional (seventh) warped
dimension, with the standard model fields confined to a 5-brane.  We
adapt and generalize the five-dimensional model of
Ref.~\cite{Grossman:2000ra}, and examine its consequences for neutrino
masses and mixing angles.  In Section 2 we present a simple effective
theory, involving a single scalar field in the seven-dimensional bulk,
that couples to the $\CN$ fields.  We then discuss, in Section 3, the
global gravitational anomalies in this context and argue that there is
no additional constraint on the number of $\CN$ fields (contrary to
claims made in the literature \cite{Grossman:2000ra, Kitano:2000wr}).

We examine the neutrino zero-modes in Section 4, where we derive their
mass matrix in terms of the wave functions of the $\CN$ zero-modes at
the standard model 5-brane.  In Section 5 we derive the shape of the
scalar vacuum expectation value (VEV), and then find the profiles of
the $\CN$ zero-modes.  The effect of the scalar VEV is to
concentrate the $\CN$ zero-modes near
the brane opposite to the standard model one,
so that the resultant
four-dimensional neutrino mass matrix is exponentially suppressed
relative to the weak scale.  We then show that the mass hierarchy
between the different flavors is limited by the presence of
higher-dimension operators, which have a flavor mixing effect.  This
is an important result, especially in view of the often stated
existence of a large exponential hierarchy induced between the masses
of different flavors, whenever the two chiralities are localized
at separate branes \cite{Grossman:2000ra, Gherghetta:2000qt}.

The implication, discussed in Section 6, is that the ratio between
the mass scales associated with the atmospheric and solar neutrinos
is not expected to be larger than one to two orders of magnitude.
Therefore, the large mixing angle MSW solution to the solar
neutrino problem is a consequence of this model.  We discuss the
neutrino mixing angles, as well as mass eigenvalues, and show that
values compatible with all current neutrino oscillation data,
except for the LSND result, emerge naturally.  The energy scales
associated with the seventh dimension, as well as the universal six
dimensions, are such that the singlet-neutrino Kaluza-Klein (KK) modes
are too heavy to play a direct role in the observed neutrino
oscillations. Similarly, the constraints from astrophysics
\cite{Barbieri:2000mg} or cosmology \cite{Goh:2001uc} on the mass
of the KK neutrinos are not relevant here.
In Section 7, we discuss the relations among various parameters in
the model and draw some conclusions about the expected mass scales
that characterize it. In Section 8, we summarize the essential
ingredients of our model leading to a viable neutrino mass spectrum,
and emphasize that the mechanism applies to a more general
class of higher-dimensional models.

%%%%%%%%%%%%%%%%%%%%%%%%%%%%%%%%%%%%%%%
\section{A Warped Seventh Dimension}
 \setcounter{equation}{0}

The seven-dimensional gauge-singlet fermions are Dirac spinors with eight
components, denoted by $\CN^i(x^M)$, where $i= 1,2,3$ labels the
generations.  The spacetime coordinates, $x^M$ with $M = 0, 1,..., 6$,
are labeled as follows: $x^0, x^1, x^2, x^3$ for the ordinary
spacetime, $x^4, x^5$ for the two additional universal dimensions, and
$x^6\equiv z$ for the dimension inaccessible to the standard model
fields.  We use the following conventions: capitals $M,N \ldots$ (from
the middle of the alphabet) denote the seven coordinate indices in a
curved background, while capitals $A,B,\ldots$ (from the beginning of
the alphabet) denote the corresponding local Lorentz indices.  We also
use lower case greek letters $\alpha,\mu,\ldots$ to refer to the
coordinate indices, and lower case latin letters, $a,m,\ldots$, to
refer to the Lorentz indices along the flat universal dimensions.

The $(T^2/Z_4)\times (S^1/Z_2)$ orbifold compactification projects out
the unwanted zero-modes, and restricts the coordinates to $0 \le x^4,
x^5 \le \pi R_u$ and $0 \le z \le \pi r_c$.  The six-dimensional
standard model fields are localized at $z=\pi r_c$, while the
gauge-singlet fields propagate in the whole bulk.

The most general metric consistent with six-dimensional Poincar\'{e}
invariance is diagonal, and warped in the $z$ direction.  However, the
compactification of the two universal extra dimensions on the
$T^2/Z_4$ orbifold breaks six-dimensional Poincar\'{e} invariance, and
in general leads to a warp factor for $x^4, x^5$ different from the
warp factor for the familiar uncompactified dimensions.  For example,
we expect contributions to the stress-energy tensor, due to the
Casimir energy of bulk fields, that do not respect the six-dimensional
Lorentz invariance.  However, we will assume for simplicity that these
differences can be neglected.\footnote{A complete solution that
incorporates the gravitational backreaction of Casimir energies or
other effects arising from the compactification of the two universal
dimensions would involve the specification of a radius stabilization
mechanism.  We leave such a study for future work.} Our main
conclusions do not change if we allow for the more general
possibility that the warp factors for the uncompactified and
compactified universal dimensions are different.  Therefore, we
consider a diagonal metric $G_{MN}$ that is warped in the $z$
direction, corresponding to a line element
\bear
\label{lineelement}
d s^2 & = & G_{MN} \, d x^M d x^N \nonumber \\ [3mm]
& = & w^2(z)
\eta_{\mu\nu} d x^\mu d x^\nu
- d z^2 ~,
\eear
where $\mu, \nu = 0, 1,..., 5$, and $\eta = {\rm diag}(+1, -1, ...,
-1)$ is the six-dimensional Minkowski metric.

Starting in Section 5 we will take the warp factor to have the form
\cite{Randall:1999ee},
\be
w(z)= e^{k (\pi r_c - z)}~,
\label{form}
\ee
which is a good approximation whenever the dominant contribution to
the bulk stress-energy tensor is due to a bulk cosmological
constant. This normalization is chosen so that $w(\pi r_c) = 1$,
which facilitates the physical interpretation at the standard model
brane, located at $z=\pi r_c$. In particular, this choice implies
that the coordinate radius $R_u$ is the \textit{proper} radius of
the universal extra dimensions as measured by standard model
probes. For the AdS metric defined by Eqs.~(\ref{lineelement}) and
(\ref{form}), the Riemann curvature tensor is
$R_{\lambda\rho\sigma\nu} = - k^2
(g_{\sigma\rho}g_{\lambda\nu}-g_{\nu\rho}g_{\lambda\sigma})$, using
the sign conventions of \cite{weinberg}.

So far we have introduced three mass parameters: the inverse
coordinate radius $1/R_u$ of the universal extra dimensions
(associated with $T^2/Z_4$), the inverse radius $1/r_c$ of the
dimension accessible only to neutral fields, and $k$. They are all
taken to be below the fundamental seven-dimensional mass scale
$M_*$, which, as will be discussed in Section 7, is in the TeV
range. We will see in Section 4 that, with the normalization $w(\pi
r_c) = 1$, the mass scale for the standard model KK modes is set by
$R_u^{-1}$, and thus satisfies the bound $R_u^{-1} \gae 0.5$ TeV,
imposed by the electroweak data \cite{Appelquist:2000nn}.
%We will see in Section 4 that $(w(\pi r_c) R_u)^{-1}$ sets the mass
%scale for the standard model KK modes, and thus satisfies the bound
%$(w(\pi r_c) R_u)^{-1} \gae 0.5$ TeV, imposed by the electroweak
%data \cite{Appelquist:2000nn}.
The radius $r_c$ of the dimension accessible only to neutral fields
is rather loosely constrained by searches for new long-range
forces.

In addition to the $\CN^i(x^M)$ fermions and the graviton, other
fields that are singlets under the standard model gauge group could be
present in the warped extra dimension.  To describe naturally small
neutrino masses it is sufficient to include a single real scalar
$\varphi$ with the dynamics described in the framework of effective
field theory.
This scalar is thus an effective degree of freedom, and could well
represent a composite structure, with the compositeness
becoming evident at scales of order $M_*$ and above.
The seven-dimensional (effective) action, invariant under both general
coordinate and local Lorentz transformations, is then given by
\bear
\int d^7 x && \hspace{-1.5em} \left\{
\sqrt{G} \left[ \frac{i}{2} \left(
\overline{\CN}^i \Gamma^A {e_{\! A}}^{\! M}\hat{D}_M \CN^i - {\rm h.c.} \right)
+ \frac{1}{2} G^{MN} \partial_M \varphi \partial_N \varphi
- V_{\varphi, \CN} \right]  \right.
\nonumber \\ [.6em]
&& \left.
\mbox{}+ \delta(z- \pi r_c ) \, \sqrt{-g} \,  {\cal L}_{\rm SM}  \right\} ~,
\label{action}
\eear
where the first two terms are kinetic terms in the warped spacetime
and the last two terms describe the bulk interactions of the $\CN$ and
$\varphi$ fields, and the six-dimensional standard model.  Here
${e_{\! A}}^{\! M}$ is the inverse vielbein, $G$ is the determinant of
the seven-dimensional metric, with $\sqrt{G} = w^6(z)$, and $g$ is the
determinant of the six-dimensional induced metric, with $\sqrt{-g} =
w^6(\pi r_c)$.  The $\Gamma^A$ are the anti-commuting matrices in
seven-dimensional Minkowski space: the gamma matrices of six-dimensional
Minkowski space along with $\Gamma^6 = i \Gamma_7$, where $\Gamma_7 =
\Gamma^0...\Gamma^5$ defines six-dimensional chirality via $\CN_\pm =
\frac{1}{2}(1\pm \Gamma_7) \CN$.  The fermion covariant derivative in
Eq.~(\ref{action}), associated with the diagonal metric $G_{MN}$, is
\be
{e_{\! A}}^{\! M} \hat{D}_M \CN = \left\{ \ba{cl}
w^{-1}(z)\left[\, \partial_\alpha + i {\delta^{a}}_{\! \alpha}\,
\Gamma_a \Gamma_7 \,(dw/dz)/2 \,\right]
\CN & , \; A = \alpha = 0, 1,..., 5  ~ \\ [0.8em]
\partial \CN/ \partial z   & , \; A = 6 ~. \ea\right.
\ee

The bulk interactions preserve the orbifold $Z_2$ symmetry, defined
such that $\CN^i_-$ and $\varphi$ are odd, while $\CN^i_+$ are even.
They may be organized into a tower of operators of increasing mass
dimension:
\be
V_{\varphi, \CN} = \hbox{} - \Lambda
- \frac{1}{2} M_\varphi^2 \varphi^2 +
\frac{\lambda_\varphi}{4 M_*^3} \varphi^4
+ \left( \frac{h_{ij}}{M_*^{3/2}} \varphi -
\frac{\bar{h}_{ij}}{M_*^{7/2}} \hat{D}^{M} \partial_M \varphi \right)
\overline{\CN^i} \CN^j  + ... ,
\label{lagn}
\ee
where $\Lambda$ is a bulk cosmological constant that needs to be
fine-tuned in order to keep flat the four dimensional sections.  At
the classical level this involves tuning $\Lambda$ against possible
brane tension terms as well as the vacuum energy stored in the
$\varphi$ VEV. The parameter $\lambda_\varphi$ is real, and $h$,
$\bar{h}$ are hermitian matrices.  All are dimensionless. The
mass-square in the second term of $V_{\varphi,
\CN}$ is chosen to satisfy $M_\varphi^2 > 0$, so that $\varphi$ has
a nonzero VEV.
Both $M_\varphi$ and $k$ are
taken to be well below $M_*$ to justify the use of effective field
theory for exploring the vacuum properties of $\varphi$.
By a flavor transformation it is possible to
diagonalize the first term inside the parenthesis, so that we can
use a basis where
\be
h_{ij} = h_i \delta_{ij} ~,
\ee
with $h_i$ real and positive. The other terms involving $\CN^i$ are
in general flavor non-diagonal.

The six-dimensional standard model Lagrangian, ${\cal L}_{\rm SM}$,
localized at $z=\pi r_c$, includes the kinetic terms for the lepton
and Higgs doublets and the Yukawa interactions of $\CN^i$:
\be
{\cal L}_{\rm SM} \supset i \overline{\CL}_-^i \Gamma^a {e_{a}}^{\!
\alpha} D_\alpha \CL_-^i + g^{\alpha\beta} D_\alpha \CH^\dagger
D_\beta \CH -  \left(
\frac{\lambda_\CN^{ij}}{M_*^{3/2}} \overline{\CL}_-^i \CN_+^j
i\sigma_2 \CH^* + {\rm h.c.} \right) ~,
\label{LSM}
\ee
where the induced (inverse) metric and vielbein at the standard model
5-brane are given by
\bear
g^{\alpha\beta} & = & w^{-2}(\pi r_{c}) \, \eta^{\alpha\beta} ~,
\nonumber \\ [.5em]
{e_{a}}^{\! \alpha} & = & w^{-1}(\pi r_{c})  \, {\delta_{a}}^{\! \alpha}
 ~, \;\;  \alpha = 0, 1,..., 5  ~.
\eear
In Eq.~(\ref{LSM}), $D_\alpha$ are the gauge covariant derivatives,
and the Yukawa couplings are again dimensionless.  Note that the
four-component field $\CL_-$ has mass dimension $+5/2$ while the
$\CN_+$ field (also four-component), being defined in seven
dimensions, has mass dimension $+3$.

Before proceeding with the analysis of the neutrino masses, we
discuss the consistency of the seven-dimensional theory.

%%%%%%%%%%%%%%%%%%%%%%%%%%%%%%%%%%%%%%%%
\section{Gravitational anomalies}
\setcounter{equation}{0}

We next show that the seven-dimensional model described in the
previous section is anomaly free.  The reader interested mostly in
neutrino phenomenology may wish to move directly to Section 4.

The seventh dimension is compactified on a
$S^{1}/Z_{2}$ orbifold and the six-dimensional standard model is
localized on a 5-brane at one of the two fixed points, while the
three singlet neutrino fields propagate in the bulk.  It was shown in
\cite{Dobrescu:2001ae} that if all fields were six-dimensional, the
resulting theory would be free of gauge and gravitational anomalies,
both \textit{local} and \textit{global}.  Letting the neutrino fields
propagate in a seventh dimension amounts to adding three infinite
towers of KK fields to this theory.  Since all gauge fields
are localized at the orbifold fixed points, allowing the singlets to
propagate in more dimensions cannot introduce any gauge anomalies.
Gravity, however, propagates in the bulk and one must consider whether
all gravitational anomalies cancel.  When coupling fermions to
gravity, there can be two types of anomalies: those associated with
general coordinate transformations and those associated with local
Lorentz transformations.  For each of these cases one must distinguish
between \textit{local} and \textit{global} anomalies.\footnote{Note
that the word ``local" can have two different meanings.  In the
context of ``local Lorentz" transformations it means that the Lorentz
group has been gauged, the standard usage in gauge theories.  There
can be, however, local Lorentz transformations that are continuously
connected to the identity as well as local Lorentz transformations
that are not.  It is customary to refer to the transformations of the
first kind as \textit{local} and to those of the second kind as
\textit{global}. We use \textit{italic fonts} whenever we want
to emphasize the distinction between the transformations that are
continuously connected to the identity and those that are not.} We
analyze first the case of \textit{local} gravitational anomalies.
After showing that there are none we turn to the more subtle issue of
\textit{global} gravitational anomalies.

%%%%%%%%%%%%%%%%%%%%%%%%%%%%%%%%%%%%%%%%
\subsection{\textit{Local} gravitational anomalies}

A noninvariance of the effective fermion action under local Lorentz
transformations would imply that the corresponding stress-energy tensor
$T^{M N}$ is not symmetric.  This would be
incompatible with general covariance and the conservation law %equation
$\nabla_{M}T^{M N}=0$.  Thus, in the presence of local Lorentz
anomalies, either general covariance is broken or $T^{M N}$ is not
conserved.  Anomalies associated with general coordinate
transformations, on the other hand, lead directly to $\nabla_{M}T^{M
N} \neq 0$.  In either case, the theory that results when gravity
becomes dynamical is inconsistent, and it is necessary to ensure that
all gravitational anomalies cancel.  However, the conditions %that are
derived from the requirement of anomaly cancellation for both kinds of
transformations are not independent.  %Indeed, a
At least in the case of
\textit{local} anomalies, it is possible to shift the anomalies of one
kind into anomalies of the other kind by adding suitable local terms to
the vacuum functional \cite{Bardeen:1984pm,Alvarez-Gaume:1984dr}.
Thus, we may concentrate only on general coordinate transformations.

%In the case of gauge theories in three or five dimensions, any
%{\textit local} gauge noninvariance can always be canceled by a bulk
%gauge Chern--Simons term \cite{Callan:sa, Arkani-Hamed:2001is},
%.  The same is true for some orbifold theories,
In the case of gauge theories in three or five dimensions, any
local gauge noninvariance, which is necessarily localized at
orbifold boundary points, can always be cancelled by a bulk
Chern-Simons term \cite{Callan:sa, Arkani-Hamed:2001is}, provided
the anomalies in the lower dimensional effective theory
vanish.\footnote{The mathematical relation between Chern-Simons
forms in odd dimensions and anomalies in even dimensions was
discussed in \cite{Faddeev:iz,Alvarez-Gaume:1984nf}.
%These
%references, however, did not consider the field theory orbifolds
%that interest us here.
}
We now argue that this is also the case for
\textit{local} general coordinate anomalies in seven
dimensions.\footnote{In five dimensions there are no {\textit
local} gravitational anomalies: the triangle diagrams always
vanish.} We follow the argument given in \cite{Arkani-Hamed:2001is}
for the spin-1 case.  The idea is to calculate the one-loop
contributions to the covariant divergence of the seven-dimensional
stress-energy tensor in the six-dimensional effective theory.  If
we regularize in such a way as to produce the covariant form of the
anomaly, it is possible to perform the calculation in any
convenient gauge.

For the analysis of anomalies,
it is sufficient to consider small fluctuations about a flat background
\be
\label{lineelementanomaly}
d s^2 = \left[\eta_{\alpha\beta} + h_{\alpha\beta}(x,z) \right]
d x^\alpha d x^\beta - d z^2 ~,
\ee
where $h_{\mu\nu} \ll 1$.  In Eq.~(\ref{lineelementanomaly}) we
took advantage of the gauge freedom\footnote{The invariance of the
line element Eq.~(\ref{lineelementanomaly}) under the reflection $z
\rightarrow
-z$ requires $G_{\alpha\beta}$ and $G_{zz}$ to be even, while
$G_{\mu z}$ should be odd. For consistency, the infinitesimal
parameters of a general coordinate transformation $\zeta^\alpha$
($\zeta^z$) should be even (odd). Although these boundary
conditions do not allow the zero mode of $G_{zz}$ to be gauged
away, this ``radion" mode has vectorlike
couplings in our theory so that it does not contribute to the
anomaly, and we do not include it here.} to set $G_{\mu z}=0$ and $G_{zz}=-1$.
We also choose the vielbein as follows:
$e_{a \alpha} = (\eta_{a \alpha} + \frac{1}{2}h_{a \alpha})$,
$e_{a z} = e_{z \alpha} = 0$
and $e_{z z} = 1$.  The fact that we take a symmetric vielbein
means that the stress-energy tensor is symmetric. (In
this gauge the absence of local Lorentz anomalies is explicit.)
%General coordinate anomalies can be read from the divergence of the
%stress-energy tensor.

The  action for a fermion $\Psi$ in the background
Eq.~(\ref{lineelementanomaly}) becomes
\bear
\label{actionanomaly}
S &=& \frac{i}{2} \int d^{7}x \, e \overline{\Psi}
\Gamma^A {e_{\! A}}^{\! M}\hat{D}_M \Psi + {\rm h.c.}
\nonumber \\
&=& \int d^{7}x \, \left[ i \, \overline{\Psi}
\Gamma^\alpha \partial_\alpha \Psi + \frac{i}{2}\, \overline{\Psi}
\Gamma^6 \!\! \stackrel{\leftrightarrow}{\raisebox{0.1mm}{$\partial$}}_{\!z}
\!\!\Psi - h_{\alpha\beta} T^{\alpha\beta} +
{\cal O}\left(h^{2}\right) \right]~,
\eear
where
\be
\label{sevendT}
T^{MN} = \frac{i}{4} \left[
\overline{\Psi} \Gamma^{(\! M} \!\!
\stackrel{\leftrightarrow}{\raisebox{0.1mm}{$\partial$}}^
{\raisebox{-2.28mm}{\scriptsize{$N\!)$}}} \!\! \Psi -
\eta^{MN} \left( \overline{\Psi} \Gamma^{\mu} \!\!
\stackrel{\leftrightarrow}{\raisebox{0.1mm}{$\partial$}}_\mu \!\! \Psi
+ \overline{\Psi} \Gamma^{6} \!\!
\stackrel{\leftrightarrow}{\raisebox{0.1mm}{$\partial$}}_z \!\! \Psi
\right) \right] ~,
\ee
and all components with an index along the seventh dimension vanish.
Here we use the notation
$\overline{\Psi}\!\!\stackrel{\leftrightarrow}{\raisebox{0.1mm}{$\partial$}}
\!\!\Psi \equiv \overline{\Psi}\partial \Psi - (\partial
\overline{\Psi})\Psi$, and in the second line of
Eq.~(\ref{actionanomaly}) as well as in Eq.~(\ref{sevendT}) it is
understood that all indices are raised and lowered with the
Minkowski metric $\eta_{\alpha\beta}$.

We now expand the fermion fields in KK modes:
\be
\label{modeexpansion}
\Psi_{\pm}(x,z) = \sum_{n} \psi_{\pm}^{(n)}(x) \xi^{\pm}_{n}(z)~,
\ee
where $\psi_{\pm}(x,z) = \frac{1}{2} (1 \pm \Gamma_{7})\psi(x,z)$, with
$\Gamma_{7} = - i \Gamma^{6}$.  The KK wavefunctions
$\xi^{\pm}_{n}(z)$, which can be taken real, are solutions to
\be
\partial_{z} \xi^{\pm}_{n} = \pm m_{n} \xi^{\mp}_{n}~,
\ee
where $\xi^{+}_{n}(z)$ and $\xi^{-}_{n}(z)$
satisfy Neumann and Dirichlet boundary conditions, respectively,
and are normalized as
\be
\label{normandcompleteness}
\int_{0}^{\pi r_{c}}\!dz \, \xi^{\pm}_{n}(z)
\xi^{\pm}_{n'}(z) =  \delta_{n n'} ~.
\ee

%\bear
%\label{normandcompleteness}
%\int_{0}^{\pi r_{c}}\!dz \xi^{\pm}_{n}(z) ~,
%\xi^{\pm}_{n'}(z) & = & \delta_{n n'} \nonumber \\ [0.9em]
%\sum_{n} \xi^{\pm}_{n}(z)\xi^{\pm}_{n}(z') & = & \delta(z-z') ~.
%\eear
%where the completeness relations are understood to hold in the spaces
%of functions satisfying Neumann and Dirichlet boundary conditions for
%$+$ and $-$ respectively.

The result of replacing the mode expansion
Eq.~(\ref{modeexpansion}) in the action Eq.~(\ref{actionanomaly})
is
\be
\label{actionanomalyKK}
S = \sum_{n} \int \! d^{6}x \left[ \overline{\psi}^{(n)}
\left(i \Gamma^{\alpha} \partial_{\alpha} - m_{n} \right)
\psi^{(n)} - \sum_{n'} h^{\alpha\beta \pm}_{n n'} \,
T_{\alpha\beta \pm}^{(n,n')} + {\cal O}\left(h^{2}\right)
\right] ~,
\ee
where
\bear
T_{\alpha\beta \pm}^{(n,n')} = \frac{1}{4} \left[
i \, \overline{\psi}_{\pm}^{(n)} \Gamma_{(\!\alpha} \!\!
\stackrel{\leftrightarrow}{\raisebox{0.1mm}{$\partial$}}_{\beta\!)}
\psi_{\pm}^{(n')} - \eta_{\alpha\beta} \left( i \,
\overline{\psi}_{\pm}^{(n)} \Gamma^{\mu} \!\!
\stackrel{\leftrightarrow}{\raisebox{0.1mm}{$\partial$}}_\mu \!\!
\psi_{\pm}^{(n')} - m_{n}\overline{\psi}_{\mp}^{(n)} \psi_{\pm}^{(n')}
-m_{n'}\
\overline{\psi}_{\pm}^{(n)} \psi_{\mp}^{(n')} \right) \right] ~,
\nonumber
\eear
and
\be
h^{\alpha\beta \pm}_{n n'}(x) = \int_{0}^{\pi r_{c}}\!dz
\xi^{\pm}_{n}(z) \xi^{\pm}_{n'}(z) h^{\alpha\beta}(x,z) ~.
\ee
The action Eq.~(\ref{actionanomalyKK}) corresponds to the
six-dimensional theory of an infinite number of fermion fields that
couple (chirally) to background fields $h^{\alpha\beta \pm}_{n n'}$
with standard gravitational couplings.  Note that when
$h_{\alpha\beta}$ is $z$-independent, the resulting couplings are
vectorlike (except for those of the zero-mode fermion) due to the
normalization condition Eq.~(\ref{normandcompleteness}), which is
the same for both chiralities.

One can calculate now the relevant square diagram
\cite{Alvarez-Gaume:1983ig} with one insertion of the operator
Eq.~(\ref{sevendT}), with the seven-dimensional fermions replaced
by their KK mode expansions.  Performing then the same
manipulations as in \cite{Arkani-Hamed:2001is}, and adding the
contribution of a brane fermion (in our model these are the
electrically neutral component of the ${\cal L}_-^i$, while the
corresponding bulk fermions, labeled generically in this section by
$\Psi$, are the $\CN^i$) to compensate for the zero-mode projected
out by the orbifold boundary conditions, one can finally write
\be
\label{covariantanomaly}
\int \! d^{6}x \zeta_M \nabla_N T^{MN} =
\frac{1}{2} \left[\delta(z-\pi r_{c}) - \delta(z)\right] \frac{1}{4\pi^{4}}
\int \left\{ \frac{1}{288} \mathrm{Tr}\left[v_{\zeta} R\right]
\mathrm{Tr}\left[R^{2}\right]  +
\frac{1}{360} \mathrm{Tr}\left[v_{\zeta} R^{3}\right]
\right\}~.
\ee
Here we used a compact differential form notation:
${R^{\alpha}}_{\!\!\beta} = \frac{1}{2} {R^{\alpha}}_{\!\!\beta\mu\nu}
\, dx^{\mu} \wedge dx^{\nu}$, where
${R^{\alpha}}_{\!\!\beta\mu\nu}(x,z)$ is the Riemann tensor
calculated from the background Eq.~(\ref{lineelementanomaly}), but
with the indices running only from $0$ to $5$.  The traces are
taken over the indices that are not saturated by differentials.
Also, ${(v_{\zeta})^{\alpha}}_{\!\!\beta} = \partial_{\beta}
\zeta^{\alpha}$, where the $\zeta^{\alpha}(x,z)$ can be thought of as the
infinitesimal parameters of a general coordinate transformation.
Due to general covariance, the result Eq.~(\ref{covariantanomaly})
holds in any gauge. In addition, we are allowed to replace the
Riemann tensor in Eq.~(\ref{covariantanomaly}) by its exact,
nonlinear expression, so that the final result holds in an
arbitrary background.

The covariant anomaly given in Eq.~(\ref{covariantanomaly}) does
not satisfy the Wess-Zumino consistency conditions and therefore
cannot be obtained from the general coordinate variation of a
functional of the metric.  There is a standard procedure to obtain
the consistent anomaly by adding local terms to the stress-energy
tensor \cite{Bardeen:1984pm}. The resulting anomaly
$Q_{6}^{1}(v_{\zeta},\Gamma,R) [\delta(z-\pi r_{c}) -
\delta(z)]/2$, where ${\Gamma^{\alpha}}_{\!\!\beta}
= {\Gamma^{\alpha}}_{\!\!\beta \mu} dx^{\mu}$ is the connection
1-form, is related to the variation of a 7-form $Q_{7}(\Gamma,R)$
that can be added to the seven-dimensional action by
\be
\label{consistentanomaly}
\delta_{v_{\zeta}} \int \! Q_{7}(\Gamma,R) =
\int \! d Q_{6}^{1}(v_{\zeta},\Gamma,R)~.
\ee
The Chern-Simons secondary characteristic class $Q_{7}(\Gamma,R)$
involves traces over all seven dimensional
indices.\footnote{$Q_{7}(\Gamma,R)$ satisfies $d Q_{7}(\Gamma,R) =
\frac{1}{(2\pi)^{4}} \left\{\frac{1}{288} \left(\mathrm{Tr}
\left[R^{2}\right]\right)^{2} + \frac{1}{360} \mathrm{Tr}
\left[R^{4}\right] \right\}$, where $R$ is the seven-dimensional Riemann
curvature two-form. For a pedagogical exposition see
\cite{Alvarez-Gaume:1984dr}.
%The relation between the parity
%anomaly in odd dimensions \cite{Redlich:1983kn} and
%even-dimensional anomalies has been investigated in
%\cite{Alvarez-Gaume:1984nf}.
} However, only the six-dimensional components contribute to the
right hand side of Eq.~(\ref{consistentanomaly}) due to the
orbifold boundary conditions on the metric and on the infinitesimal
parameters $\zeta^M$.

The modified (consistent) form of Eq.~(\ref{covariantanomaly})
matches precisely with Eq.~(\ref{consistentanomaly}).
%which shows that the general coordinate variation of the seven-dimensional
%Chern--Simons term $Q_{7}(\Gamma,R)$ is a total derivative.  Indeed,
We note that $Q_{7}(\Gamma,R)$ is odd under parity (defined as
reflection through the $z = 0$ hyperplane).  Therefore, if we
define the orbifold theory by starting from a compactification on
the circle $S^{1}$, the coefficient of the Chern--Simons term must
change sign when crossing $z = 0$ (and $z = \pi r_{c}$), so that
the theory is invariant under the reflection that is used in the
orbifold projection.  Due to this discontinuity, the gauge
variation of such a term gives rise to delta-function singularities
as in Eq.~(\ref{covariantanomaly}). Alternatively, we can think of
the $S^{1}/Z_{2}$ orbifold as a compactification on an interval
(the half circle) with certain boundary conditions imposed at the
end points.  In this picture the coefficient of the Chern--Simons
term is constant and the compensating anomaly comes from the
boundary contributions.  In either picture, it is clearly possible
to cancel the noninvariance of the original fermion effective
action by including the seven-dimensional Chern--Simons form
$Q_{7}(\Gamma,R)$. We therefore assume that this Chern--Simons term
is present so that the vacuum functional is invariant under
\textit{local} coordinate transformations.

%%%%%%%%%%%%%%%%%%%%%%%%%%%%%%%%%%%%%%%%
\subsection{\textit{Global} gravitational anomalies}

There still remains the question of \textit{global} general
coordinate/local Lorentz transformations
\cite{Alvarez-Gaume:1983ig,Witten:1985xe}.  If there are
diffeomorphisms not continuously connected to the identity, the
previous analysis is not enough to ensure that the theory is invariant
under such transformations.  We phrase the following analysis in terms
of general coordinate transformations, but the same arguments apply
for the case of local Lorentz transformations.

If $W(G)$ denotes the fermion determinant in the presence of a
background metric $G$, we have in general
\be
\label{anomalousphase}
\frac{W(G^{\rho})}{W(G)} = e^{i \delta_{[\rho]}}~,
\ee
where $G^{\rho}$ denotes the metric obtained from $G$ under a
representative $\rho$ of one of the disconnected classes of
diffeomorphisms, and $\delta_{[\rho]}$ is a phase that depends on the
class to which $\rho$ belongs.\footnote{We note that the group of
disconnected diffeomorphisms of the $n$-sphere $S^{n}$ is trivial for
$n\leq 5$.  This follows from the absence of ``exotic''
$(n+1)$-spheres for $n+1\leq 6$ \cite{Milnor:1965}.  There are
therefore no constraints on the number of bulk neutrinos in the
popular five-dimensional $S^{1}/Z_{2}$ orbifolds from \textit{global}
anomaly considerations.  For higher dimensional theories the situation
is not as straightforward, as indicated by the
fact that there are 28 disconnected components on $S^6$ and two on
$S^7$.  See \cite{Kervaire:1963} for some other higher dimensional
cases.} We are specifically interested in the diffeomorphism classes
of $\mathcal{M} = S^{4} \times (T^{2}/Z_{4}) \times (S^{1}/Z_{2})$. The
key observation is that $S^{1}/Z_{2}$ is diffeomorphic to an interval
and any diffeomorphism of $\mathcal{M}$ onto $\mathcal{M}$ can be
continuously deformed into one which is trivial in the interval
$S^{1}/Z_{2}$:
\be
\label{specialTrans}
x'^{\alpha} = f^{\alpha}(x) \hspace{1cm} z' = z~, \nonumber
\ee
where $x^{\alpha}$ denote the coordinates in $S^{4}
\times (T^{2}/Z_{4})$.  We can therefore restrict attention to coordinate
transformations of the type (\ref{specialTrans}). It is then
convenient to perform a KK decomposition for all fields, including
the background metric, and analyze the resulting six-dimensional
theory. Regarding the background, we note that under the limited
class of diffeomorphisms (\ref{specialTrans}) the affine connection
transforms as
\be
\label{connectiontransf}
\Gamma'^{\lambda}_{\alpha \beta} =
\frac{\partial{x'^{\lambda}}}{\partial{x^{\rho}}}
\frac{\partial{x^{\tau}}}{\partial{x'^{\alpha}}}
\frac{\partial{x^{\sigma}}}{\partial{x'^{\beta}}} \Gamma^{\rho}_{\tau \sigma} +
\frac{\partial{x'^{\lambda}}}{\partial{x^{\rho}}}
\frac{\partial^{2}{x^{\rho}}}{\partial{x'^{\alpha}}\partial{x'^{\beta}}}~,
\ee
while all other components transform simply as tensors (the second
term in Eq.~(\ref{connectiontransf}) would vanish if any of the
indices $\alpha, \beta, \lambda$ were along the seventh dimension
parameterized by $z$).  Furthermore, if we perform a suitable KK
mode expansion, only the zero mode of $\Gamma^{\lambda}_{\alpha
\beta}$ is affected by the second term in
Eq.~(\ref{connectiontransf}); all other KK modes are true tensors
under (\ref{specialTrans}).  (When referring to the gravitational
background field, a zero mode is \textit{defined} to be independent
of the higher-dimensional coordinate.)

We next argue that if there are no \textit{global} anomalies in a
purely zero-mode gravitational background, then there are no
\textit{global} anomalies even in the presence of the higher
gravity KK modes.  The reason for this is that the group of
disconnected diffeomorphisms is finite, at least for the case of
the $n$-sphere $S^{n}$ \cite{Kervaire:1963}. It follows that for
any element $\rho$ in the group there exists a (smallest) integer
$N$ such that $\rho^{N}$ is the identity element, and therefore the
phase in Eq.~(\ref{anomalousphase}) associated with $\rho$ must be
an integer multiple of $2 \pi/N$. If this phase vanishes when the
higher gravity KK modes are turned off, and we turn them on
smoothly, the phase must remain zero, unless it changes
discontinuously. Since the higher gravity KK modes are just like
background ``matter'' fields in the appropriate tensor
representation of (\ref{specialTrans}), we consider this very
unlikely.

Now, in a zero-mode gravity background, the theory in question is
just the six-dimensional standard model with the addition of three
infinite towers of massive neutrino KK modes, which have
\textit{vectorlike} couplings to the background gravity field.  We
also note that the Chern--Simons term that is needed to cancel the
\textit{local} anomalies is invariant under (\ref{specialTrans}) when
the higher gravity KK modes are turned off. Therefore, the fermion
effective action in a zero-mode gravity background is invariant
under general coordinate transformations. From the argument given
in the previous paragraph it follows that there are no
\textit{global} anomalies in an \textit{arbitrary} gravitational
background.  It is worth pointing out that in the presence of the
higher gravity KK modes, the Chern--Simons term $Q_{7}(\Gamma,R)$
is not invariant even under the restricted class of diffeomorphisms
(\ref{specialTrans}). This noninvariance must be canceled by the
rest of the terms involving the higher gravity KK modes.

We conclude that adding a seventh dimension compactified on $S^1/Z_2$
to the six-dimen-sional standard model and letting the neutrinos
propagate in the bulk, introduces neither \textit{local} nor
\textit{global} gravitational anomalies.  Turning the argument around,
we can say that the consistency constraints on the number of neutrinos
in the seven-dimensional model are the same as in the six-dimensional
standard model analyzed in \cite{Dobrescu:2001ae}, namely it is
necessary to include one singlet neutrino per generation.  From the
point of view of anomaly cancellation it is immaterial whether these
neutrinos are bulk or brane fields.

%%%%%%%%%%%%%%%%%%%%%%%%%%%%%%%%%%%%%%%%
\section{Neutrino Masses}
\setcounter{equation}{0}

We now return to the action of Eq.~(\ref{action}), which leads to
Dirac neutrino masses.  In order to study the implications for
neutrino oscillations it is sufficient to analyze the zero-mode
spectrum.

The zero modes of $\CN_{-}^{i}$ are projected out by the orbifold
boundary conditions.  The KK decomposition along the warped dimension
that includes the zero-mode $\CN$ fields is given by
\be
\CN^i_+(x^\alpha, z) = \frac{1}{\sqrt{\pi r_{c}}}
\sum_{n=0}^{\infty} \CN^{i(n)}_+(x^\alpha) \xi^{n}_i(z)~,
\label{decom}
\ee
where the index $\alpha = 0,1, ..., 5$ labels the universal
dimensions.  The $\xi^{n}_i(z)$ form a complete set of orthogonal
(dimensionless) functions on the $[0, \pi r_c]$ interval, satisfying
Neumann boundary conditions appropriate for even fields.  They are
chosen to obey the ortho-normality conditions
\be
\label{normal}
\frac{1}{\pi r_{c}}\int^{\pi r_c}_0 dz \, w^5(z)
\xi^{n*}_i(z)\xi^{n^\prime}_i(z)
= \delta_{n n^\prime}~,
\ee
which ensure the canonical normalization of the six-dimensional
kinetic terms for $\CN^{i(n)}_+$.

%It is also convenient to canonically normalize the six-dimensional kinetic
%terms of the standard model fields.  To this end we perform the
%following field redefinitions.  The quark and charged lepton fields
%are all rescaled as
%\be
%\CL_{-}(x^\alpha)  \rightarrow  w^{-5/2}(\pi r_c) \, \CL_{-}(x^\alpha) ~,
%\ee
%and the Higgs field is rescaled as
%\be
%\CH (x^\alpha) \rightarrow  w^{-2}(\pi r_c) \, \CH (x^\alpha) ~.
%\label{h-rescale}
%\ee

We now adopt a warp factor chosen to be unity at the standard model
brane, as in Eq.~(\ref{form}). All kinetic terms for the standard
model fields are then automatically canonically normalized. Keeping
the zero-modes with respect to the seventh dimension only, and
integrating over $z$, the six-dimensional effective Lagrangian [see
Eq.~(\ref{action})] is
\bear
\label{normlagr}
{\cal L}_{6D} &=& i\overline{\CL}_-^i \Gamma^\alpha D_\alpha
\CL_-^i + D_\alpha \CH^\dagger D^\alpha \CH + i
\overline{\CN}_+^{i(0)} \Gamma^\alpha \partial_\alpha \CN_+^{i(0)}
\nonumber \\ [.5em]
& & -
\left( \frac{\lambda_\CN^{ij} \,
 \xi^{0}_j(\pi r_{c}) }{
\sqrt{\pi r_{c}} \, M_*^{3/2} }
 \, \overline{\CL}_-^i \CN_+^{j(0)}
i\sigma_2 \CH^* + {\rm h.c.} \right) + ... ~,
\eear
where $\alpha = 0,1,..., 5$.  Note that we do not need to
distinguish between coordinate and Lorentz indices anymore, and all
the indices are raised and lowered with the flat metric
$\eta_{\alpha\beta}$. Eq.~(\ref{normlagr}) shows that the mass
scale for the standard model KK modes is set by the inverse
\textit{proper} radius $1/R_u$ [or $1/(w(\pi r_c) R_u)$ for an
arbitrary normalization of the warp factor.]
%[For a general
%normalization of the warp factor, which can be obtained from the
%previous case by a global rescaling of the coordinates $x^\mu$ and
%as such cannot change the physics, one would find that the scale
%for the standard model KK masses is set by the inverse proper
%radius as measured with the new induced metric. If one insists on
%defining $R_u$ as the \textit{coordinate} radius of the universal
%dimensions (for the rescaled coordinates), the scale for the KK
%masses would be given by $1/(w(\pi r_c) R_u)$.]

Integrating out the universal extra dimensions, the Dirac neutrino
mass matrix induced after electroweak symmetry breaking is
\be
\label{massmatrix}
M_{\nu}^{ij} = \frac{\lambda_\CN^{ij}\,  v_h}{\pi R_{u}M_*
\sqrt{\pi r_{c}M_* }  } \, \xi^{0}_j(\pi r_{c}) ~,
\ee
where $v_h = 174~\mbox{GeV}$ is the Higgs VEV, and the denominator
represents the square-root of the volume of the $(T^2/Z_4) \times
(S^1/Z_2)$ orbifold.  As we will see in the next section, the
neutrino mass eigenvalues are largely determined by the hierarchy
among the $\xi^0_j(\pi r_c)$, while the mixing angles are
determined by the flavor structure of the couplings
$\lambda_\CN^{ij}$ and $\bar{h}_{ij}$.

%%%%%%%%%%%%%%%%%%%%%%%%%%%%%%%%%%%%%%%%
\section{Zero-mode Profiles of Gauge-Singlet Fermions}
\setcounter{equation}{0}

In this section we derive the profiles of the neutrino zero modes,
which determine the neutrino mass matrix according to
Eq.~(\ref{massmatrix}).  These depend on the VEV of $\varphi$ and
therefore our first task is to determine the solution to the
$\varphi$ equation of motion that follows from the
$\CN$-independent part of Eqs.~(\ref{action}) and (\ref{lagn}).

%%%%%%%%%%%%%%%%%%%%%%%%%
\subsection{The bulk VEV}

We will be interested in a region of parameter space where the
$\varphi$ field VEV varies slowly in the bulk of the 7th dimension
(with the exception of two narrow regions close to the branes), so
that to a good approximation it simply gives a contribution to the
bulk cosmological constant.  Thus, we use the explicit form for the
warp factor, Eq.~(\ref{form}).

The negative mass-squared of $\varphi$ implies that a
nonzero VEV for $\varphi$ is energetically favored,
but at the same time $\varphi$ is an odd field under
the orbifold identification, and therefore its VEV
must satisfy the boundary conditions
\be
\langle \varphi(0) \rangle = \langle \varphi(\pi r_c) \rangle = 0~.
\ee
%The VEV of $\varphi$ is $z$-dependent, and is a solution to the
%classical equation of motion of $\varphi$ in the background
%(\ref{lineelement}).
In terms of the rescaled VEV,
\be
\label{deff}
f(z) = \frac{\langle \varphi(z) \rangle}{M_*^{3/2}} ~,
\ee
which has mass dimension +1, the equation of motion is
\be
\frac{d^2 f}{d z^2}
- 6 k \frac{d f}{d z} = \lambda_\varphi f^3 - M_\varphi^2 f ~,
\label{VEV1}
\ee
subject to the boundary conditions $f(0) = f(\pi r_c) = 0$.
We assume $k >0$.  In Eq.~\ref{VEV1} we have neglected
possible higher dimension operators.  This is justified as long as the
effective field theory description is valid, that is as long as $k$ and
$M_\varphi$ are well below $M_*$.
%Since $\varphi$ is an odd field under
%The orbifold identification boundary conditions are
%$f(0) = f(\pi r_c) = 0$.

Eq.~(\ref{VEV1}) describes the motion of a particle in the potential
\be
V(f)= - \frac{\lambda_\varphi }{4} f^4 + \frac{M_\varphi^2}{2} f^2~,
\ee
in the presence of an \textit{anti}-friction term proportional to $k$.
Thus, we are looking for trajectories in which the particle starts at
the bottom of the potential $f = 0$ with some initial velocity,
climbs the potential up to a certain point and then rolls down back to
$f = 0$.  The anti-friction term puts energy into the system, so
it is conceivable that for a sufficiently large $k$, no matter how
small the initial velocity, the particle will gain enough energy to
overcome the potential barrier; in this case the only solution that
satisfies the boundary conditions is the trivial one $f(z) = 0$.

We first determine the restrictions in parameter space for nontrivial
solutions to exist.  In order to do this it is convenient to set
$\tilde{f}(z) = \sqrt{\lambda_\varphi} \, M_\varphi^{-1} \, e^{-3kz}
f(z) $, so that Eq.~(\ref{VEV1}) becomes
\be
\frac{d^2 \! \tilde{f}}{d z^2}
=  M_\varphi^2 e^{6kz} \tilde{f}^3 - \left( M_\varphi^2 - 9 k^2 \right)
\tilde{f} ~.
\label{VEV2}
\ee
describing now frictionless motion in a potential whose slope
decreases with time.  For $k \geq M_\varphi/3$, the ``motion'' starts
at $z=0$ from the \textit{maximum} $\tilde{f}=0$ of a continuosly
decreasing potential, so that $\tilde{f}(z) \equiv 0$ is the only
solution satisfying $\tilde{f}(\pi r_c) = 0$.  Therefore,
Eq.~(\ref{VEV2}) can have a nontrivial solution which satisfies
$\tilde{f}(0) = \tilde{f}(\pi r_c) = 0$ only if
\be
k < \frac{M_\varphi}{3} ~.
\label{bound}
\ee
This necessary condition is not sufficient for the existence of
solutions with $\tilde{f}(0) = \tilde{f}(\pi r_c) = 0$.  Another
necessary condition can be derived as follows.  If we neglect the
first term on the right-hand-side of Eq.~(\ref{VEV2}) the fictitious
particle feels just a harmonic oscillator potential and returns to the
origin after a ``time'' $z = \pi (M_\varphi^2 - 9 k^2)^{-1/2}$.  The
effect of the neglected term is always to increase the oscillation
period.  Hence, the boundary condition $\tilde{f}(\pi r_c) = 0$
requires
\be
\frac{1}{r_c^2} < M_\varphi^2 - 9 k^2~.
\label{bound2}
\ee
In the Appendix we prove that the conditions (\ref{bound}) and
(\ref{bound2}) are also sufficient for the existence of nontrivial
solutions.  We conclude that the VEV $\langle \varphi(z) \rangle$
is non-zero in the
interval $ 0 < z < \pi r_c$ for a substantial range of values of
$M_{\varphi}$ and $k$.

\begin{figure}[t]
\begin{center}
\scalebox{1.2}[1]{\includegraphics{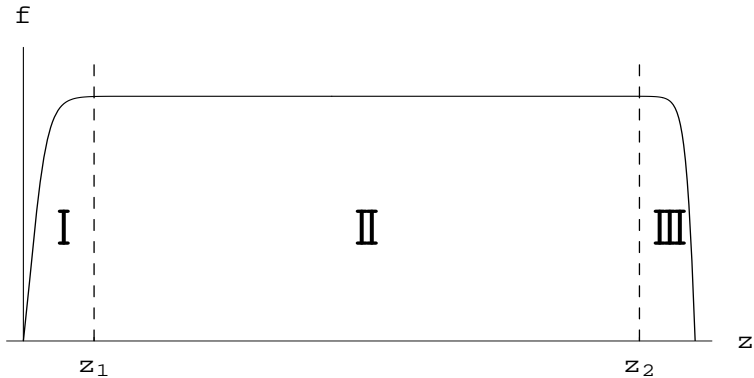}}
\par
\vskip-2.0cm{}
\end{center}
\caption{\small Scalar profile when $M_{\varphi}/k=10$, $\pi r_{c}
M_{\varphi}=70$.
The plateau is at $f \approx M_\varphi/\sqrt{\lambda_\varphi}$.
The points $z_{1}$ and $z_{2}$
define the boundaries between regions I, II and III. In the main text
we consider the case where $z_{1}/z_{2} \ll 1$.}
\label{profile}
\end{figure}

For the special case of a flat and large $z$-dimension, $k \ll
1/(\pi r_c) \ll M_\varphi$, the solution is given approximately by
\cite{Georgi:2000wb}
\be
f(z) \approx
\frac{M_\varphi}{\sqrt{\lambda_\varphi}} \,
\tanh \frac{M_\varphi z}{\sqrt{2}} \,
\tanh \frac{M_\varphi (\pi r_c - z)}{\sqrt{2}}
\, \left[ 1 + {\cal O}\left(\frac{k}{M_\varphi}\right)
+ {\cal O}\left(e^{-\pi r_c M_\varphi}\right)
\right] ~.
\ee
This solution is essentially constant except in the region of size
$\sim M_{\varphi}^{-1}$ around the endpoints.  We will in general
be interested in the parameter range in which $k$ is not
negligible, but where $1/(\pi r_{c}) \ll M_{\varphi}$.  The latter
hierarchy will need to be only one to two orders of magnitude to
explain the smallness of the neutrino masses relative to the weak
scale.  As we show in the Appendix, the solutions in this case are
qualitatively similar to the flat case $k=0$.  We show a typical
numerical solution in Fig.~\ref{profile}.

%%%%%%%%%%%%%%%%%%%%%%%%%
\subsection{Bulk fermions}

In the presence of the $\varphi$ VEV, the $\CN$ fields have
non-trivial profiles along the $z$ dimension.  The three $\CN $
zero-modes defined in Eq.~(\ref{decom}) are a solution to the set of
equations ($i, j=1,2,3$, with $i$ fixed and $j$ summed over):
\be
\label{three}
\frac{d \xi^0_i}{d z} = \left(3 k \delta_{ij} -
h_{i} \delta_{ij} f -
\frac{\bar{h}_{ij}}{M_*^{2}} f^{\prime\prime}
\right) \xi^0_j~,
\ee
where we have again neglected possible higher dimension operators
in Eq.~(\ref{lagn}).  We also set $f^{\prime\prime}=d^2 f /d z^2$
where $f$ was defined in Eq.~(\ref{deff}).  It will be useful to
factor out the leading order solution in powers of $M_\varphi/M_*$,
namely the solution in the absence of the last term in
Eq.~(\ref{three}), by defining new functions $c_i(z)$ through
\be
\label{defc}
\xi^0_i(z) = c_i(z) e^{3kz - h_i S(z)}~,
\ee
where
\be
\label{exponent}
S(z) \equiv \int_0^z d\zeta \, f(\zeta)~.
\ee
The $c_{i}$'s satisfy the following differential equations
\be
\label{eqc}
\frac{d c_i}{d z} = - \frac{\bar{h}_{ij}}{M_*^{2}}
f^{\prime\prime}  e^{(h_i-h_j) S(z)} c_j~,
\ee
which we now solve in the limit $\pi r_c M_\varphi \gg 1$.

Given the general features of the $f$ profile discussed in the
previous subsection, it is convenient to separate the analysis in the
three regions shown in Fig.~\ref{profile}.  We first note that in
region II, $f^{\prime\prime} $ is exponentially small and therefore
all the $c_{i}$'s remain essentially constant throughout it.  The
differential equations (\ref{eqc}) are nontrivial in regions I and
III. In region I the integral Eq.~(\ref{exponent}) is of order
$\lambda_{\varphi}^{-1/2}$. % unity.
%Thus, if all the $c_{i}$ at $z=0$ are of the same order then this will
%still be true for the $c_{i}(z_{1})$ when $h_i\lambda_{\varphi}$ is of order
%unity, and therefore also for the $c_{i}(z_{2})$.
In region III this integral is much larger, so that
the important features of the $c_{i}(z)$ are determined
in this region as follows: for $z_{2}<z<\pi r_{c}$ the integral
expression for $S(z)$, Eq.~(\ref{exponent}), can be replaced, to a good
approximation, by $S(\pi r_c)$, which is itself of order
$\lambda_{\varphi}^{-1/2} \pi r_c M_{\varphi}$.  Therefore, in
this region
there will be an exponential hierarchy among the various
terms on the right-hand-side of Eqs.~(\ref{eqc}),
provided $h_i\lambda_{\varphi}^{-1/2} \pi r_c M_{\varphi} \gg 1$.
Without loss of
generality we can assume the ordering $h_1 > h_2 > h_3 > 0$.  If we
keep only the leading terms, then Eqs.~(\ref{eqc}) in region III reduce
to
\bear
\label{eqc3}
\frac{d c_3}{d z} & \simeq & - \frac{\bar{h}_{33}}{M_*^{2}}
 f^{\prime\prime} \, c_3 \\ [0.5em]
\label{eqc12}
\frac{d c_i}{d z} & \simeq & - \frac{\bar{h}_{i3}}{M_*^{2}}
f^{\prime\prime} e^{(h_i-h_3) S(\pi r_c)} \, c_3 \hspace{1cm} \mbox{for}
\hspace{2mm} i = 1,2 ~.
\eear

We first solve for $\xi^0_3$.  It will be sufficient to work to
zero-th order in $M_{\varphi}/M_{*}$, so that from Eq.~(\ref{eqc3})
we have $c_3(z) = \mbox{const}$.  Imposing the normalization
condition Eq.~(\ref{normal}), and in the limit $1/(\pi r_c) \ll k <
M_\varphi/3$, where we can evaluate $S(z)$ by setting $f(z) =
M_\varphi/\sqrt{\lambda_{\varphi}}$ throughout the region of
integration, we find from Eq.~(\ref{defc}) that
\be
\label{solxi3}
\xi^0_3(\pi r_c) \simeq
\sqrt{\pi r_c \left( 2\tilde{M}_{\varphi} - k \right) } \,
e^{- \pi r_c \left(\tilde{M}_{\varphi} - k/2 \right)}~,
\ee
where only one combination of the parameters in the Lagrangian for
$\varphi$ appears,
\be
\tilde{M}_{\varphi} \equiv
\frac{h_{3} M_{\varphi}}{\sqrt{\lambda_{\varphi}} } ~.
\ee
We will see in the next section that the expression
Eq.~(\ref{solxi3}) will lead to exponential suppression of neutrino
masses provided only that $\tilde{M}_{\varphi} > k/2$.
%Note the exponential factor in Eq.~(\ref{solxi3}), arising
%from the coupling of $\CN^{i}$ to the $\varphi$ VEV.
%In the next section we will see that the smallness of the neutrino masses
%requires $\tilde{M}_{\varphi} > k/2$.

Na\"\i vely, one might think that for the other two generations
there will be a greater exponential suppression controlled by their
larger Yukawa couplings $h_{1,2}$.  We now show that this is not
the case, due to the presence of the higher dimension operators
that couple $\CN^{1,2}$ to $\CN^{3}$ in Eq.~(\ref{lagn}). To see
this we need to solve for $c_{1,2}(z)$ in region III to first order
in $M_{\varphi}^{2}/M_{*}^{2}$.  By using the zero-th order
solution for $c_{3}(z) \simeq c_{3}(\pi r_{c})$ in
Eq.~(\ref{eqc12}) we obtain
\be
c_i(z) \simeq c_i(z_2) - c_3(\pi r_c) \frac{\bar{h}_{i3}}{M_*^{2}}
e^{(h_i-h_3) S(\pi r_c)} \left[ f^\prime(z) - f^\prime(z_2)\right]~,
\hspace{5mm} \mbox{for} \hspace{2mm} i = 1,2,
\ee
from which we find
\be
\frac{\xi^0_i(\pi r_c)}{\xi^0_3(\pi r_c)}
\simeq \frac{c_i(z_2)}{c_3(\pi r_c)}e^{-(h_i-h_3) S(\pi r_c)}
- \frac{\bar{h}_{i3}}{M_*^{2}}
\left[ f^\prime (\pi r_c) - f^\prime (z_2) \right]~.
\label{xi-ratio}
\ee
But the first term here is exponentially small compared to the second term.
%we argued before that the
%ratio $c_i(z_{2})/c_3(z_{2}) \sim
%c_i(z_{2})/c_3(\pi r_c)$ is exponentially  compared with
%the exponential factor in Eq.~(\ref{xi-ratio}),
%we can neglect the first term compared to the second.
Furthermore, $f^\prime(z_2)$ is
exponentially small compared to $f^\prime(\pi r_c)$ and we finally
obtain
\be
\label{solxi12}
\frac{\xi^0_i(\pi r_c)}{\xi^0_3(\pi r_c)} \simeq
- \frac{\bar{h}_{i3}}{M_*^{2}} f^\prime(\pi r_c)~.
\ee
Because $f(z)$ varies from ${\cal O}(\lambda_\varphi^{-1/2}
M_{\varphi})$ to zero over a distance of order $1/M_{\varphi}$, it
follows that $-f^\prime(\pi r_{c}) = {\cal O}(\lambda_\varphi^{-1/2}
M_{\varphi}^2)$.  This provides the promised result.  The zero-mode
wave functions $\xi^0_{1,2}(\pi r_c)$ are suppressed compared to
$\xi^0_3(\pi r_c)$, but only by a quantity
\be
\epsilon \equiv - \frac{f^\prime(\pi r_c)}{M_{*}^{2}} \sim
{\cal O} \left(\frac{M_{\varphi}^{2}}{\sqrt{\lambda_\varphi}
M_{*}^{2}} \right) ~.
\label{epsilon}
\ee
We note that in Eq.~(\ref{xi-ratio}) the leading term in $1/M_*$ is
exponentially suppressed while the subleading term is not, so that
it dominates as long as $\bar{h}_{i3}$ is not extremely small
compared with unity.  The reader might wonder whether this signals
a breakdown in the effective theory description, which relies on
the convergence of the expansion in $1/M_*$.  There is no reason to
worry: the term suppressed by $M_*$ in Eq.~(\ref{xi-ratio}) comes
actually from the {\it leading} flavor off-diagonal operator.  All
other terms suppressed by higher powers of $M_*$ give just small
corrections to $\epsilon$.

%%%%%%%%%%%%%%%%%%%%%%%%%%%%%%%%%%%%%%%%%%%%%%%%%%
\section{Neutrino Oscillations}
\setcounter{equation}{0}

We are now equipped with all the tools necessary for analyzing the
neutrino mass spectrum and the ensuing neutrino oscillations.

%%%%%%%%%%%%%%%%%%%%%%%%%
\subsection{Neutrino mass matrix}

As we discussed in the introduction (Section 1), the neutrino
masses are of the Dirac type, the Majorana masses being forbidden
by the symmetry under rotations of the two universal extra
dimensions.  Below the electroweak scale, the effective
four-dimensional theory contains three left-handed neutrinos,
$\nu^i_L$, which are the neutral zero-modes of $\CL^i_-$, and three
right-handed neutrinos, $N_R^i$, which are the zero-modes of
$\CN^{i(0)}_+$ with respect to the two universal dimensions.  In
the weak eigenstate basis,
\be
\bar{\nu}_L M_\nu N_R + {\rm h.c.} ~,
\ee
the neutrino mass matrix derived in Eqs.~(\ref{massmatrix}), (\ref{solxi3})
and (\ref{solxi12}) is given by
\be
M_\nu = m_0 \left( \ba{ccc}
\epsilon \, \bar{h}_{13} \lambda_\CN^{11} \ & \
\epsilon \, \bar{h}_{23} \lambda_\CN^{12} \ & \
\lambda_\CN^{13}
\\ [1em]
\epsilon \, \bar{h}_{13} \lambda_\CN^{21} &
\epsilon \, \bar{h}_{23} \lambda_\CN^{22} &
\lambda_\CN^{23}
\\ [1em]
\epsilon \, \bar{h}_{13} \lambda_\CN^{31} &
\epsilon \, \bar{h}_{23} \lambda_\CN^{32} &
\lambda_\CN^{33}
\ea \right) ~,
\ee
where $\epsilon \ll 1$ is defined in Eq.~(\ref{epsilon}),
and the scale of the neutrino masses is set by
\be
m_0 \equiv v_h e^{- \pi r_c \left(\tilde{M}_{\varphi} - k/2
\right)} \left(\frac{1}{\pi R_u M_*} \right)
\left(\frac{2\tilde{M}_{\varphi} - k}{M_*} \right)^{\! 1/2}~.
\label{expo}
\ee
As we will see in Section 7, the factors in parenthesis are
expected to provide a suppression of no more than one to two orders
of magnitude, so that the neutrino mass scale is explained by the
first exponential.

The neutrino mass matrix is diagonalized by
unitary transformations:
\be
U_L^\dagger M_\nu U_R
= {\rm diag}\left( m_{\nu_1}, \, m_{\nu_2}, \, m_{\nu_3} \right) ~.
\ee
The unitary matrix describing neutrino oscillations, $U_L$, and the
physical neutrino masses, $m_{\nu_i}$, may be found by expanding in powers of
$\epsilon = {\cal O} (\lambda_\varphi^{-1/2} M_{\varphi}^{2}/M_{*}^{2})$.
The largest physical neutrino squared-mass is
\be
m_{\nu_3}^2 = m_0^2 \left( \left|\lambda_\CN^{13}\right|^2
+ \left|\lambda_\CN^{23}\right|^2 + \left|\lambda_\CN^{33}\right|^2 \right)
\left[ 1 + {\cal O}(\epsilon^2) \right] ~.
\label{nu3}
\ee
It is convenient to use the following identity, valid up to corrections of
order $\epsilon^2$:
\be
M_\nu = \left( \ba{ccc}
\tilde{l}_3 & - (l_{23}\tilde{l}_1)^* & l_{13}
\\ [1em]
0 & \sqrt{ |l_{13}|^2 + |l_{33}|^2 } & l_{23}
\\ [1em]
-\tilde{l}_1 & - (l_{23}\tilde{l}_3)^* & l_{33}
\ea \right)
\left( \ba{ccc}
\epsilon \, \tilde{m}_{11} & \epsilon \, \tilde{m}_{12} & 0
\\ [1em]
\epsilon \, \tilde{m}_{21} & \epsilon \, \tilde{m}_{22} & 0
\\ [1em]
0 & 0 & m_{\nu_3}
\ea \right)
\left( \ba{ccc}
1 & 0 & -\epsilon\theta_1
\\ [1em]
0 & 1 & -\epsilon\theta_2
\\ [1em]
\epsilon \theta_1 & \epsilon \theta_2 & 1
\ea \right) ~,
\label{sand}
\ee
where we use the following notation:
\bear
l_{ij} & = & \frac{\lambda_\CN^{ij} }{ \left( \left|\lambda_\CN^{13}\right|^2
+ \left|\lambda_\CN^{23}\right|^2 + \left|\lambda_\CN^{33}\right|^2 \right)^{1/2}}
\;\;\;\; , \;\;\;\; i,j = 1,2,3~,
\nonumber \\ [1em]
\tilde{l}_i & = & \frac{{\lambda_\CN^{i3}}^* }{ \left( \left|\lambda_\CN^{13}\right|^2
+ \left|\lambda_\CN^{33}\right|^2 \right)^{1/2}}
\;\;\;\; , \;\;\;\; i = 1,3~,
\nonumber \\ [1em]
\theta_i & = & \bar{h}_{i3} \sum_{j=1}^{3} l_{ji}^* l_{j3}^* \;\;\;\; , \;\;\;\;
i = 1,2 ~.
\eear
The first matrix that appears in Eq.~(\ref{sand}) can be shown to be unitary,
while the last matrix is unitary to leading order in $\epsilon^2$.
%and last matrices shown in Eq.~(\ref{sand}) are unitary.
%The unitarity of the first one can be shown exactly, while for the
%last matrix one can check unitarity to leading order in $\epsilon^2$.
The elements of the block-diagonal mass matrix shown in
Eq.~(\ref{sand}) are given by
\bear
\tilde{m}_{1i} & = & m_{\nu_3} \bar{h}_{i3} \left( l_{1i} \tilde{l}_3^*
- l_{3i} \tilde{l}_1^* \right) ~,
\nonumber \\ [1em]
\tilde{m}_{2i} & = & m_{\nu_3} \bar{h}_{i3} \left[
\tilde{l}_1 \left( l_{13}l_{2i} - l_{23}l_{1i} \right)
+ \tilde{l}_3 \left( l_{33}l_{2i} - l_{23}l_{3i} \right)
 \right] \;\;\;\; , \;\;\;\; i = 1,2~.
\label{mtilde}
\eear

Eq.~(\ref{sand}) shows that the $m_{\nu_1}$ and $m_{\nu_2}$
physical neutrino masses are of order $\epsilon m_{\nu_3}$, and
generically are non-degenerate. We choose $m_{\nu_1} < m_{\nu_2}$.
These can be computed straightforwardly by diagonalizing
$\tilde{m}$ (the $2\times 2$ matrix whose elements are given by
$\tilde{m}_{ij}$):
\be
\epsilon^2 \tilde{m}\tilde{m}^\dagger = V {\rm diag}
\left( m_{\nu_1}^2, \, m_{\nu_2}^2 \right)
V^\dagger~,
\ee
where $V$ is a unitary $2\times 2$ matrix. Its $V_{ij}$ elements
depend only on ratios of $\lambda_\CN^{ij}$'s
and on $|\bar{h}_{13}/\bar{h}_{23}|$. This dependence can be computed
straightforwardly using Eq.~(\ref{mtilde}), but is cumbersome and we do not
display it here.

The unitary $3\times 3$ matrix $U_L$ is then given by the product of the
first matrix on the right-hand-side of Eq.~(\ref{sand}) with
\be
\left( \ba{ccc}
V & 0
\\ [.51em]
0 & 1
\ea \right) ~. \nonumber
\label{vl}
\ee
The third column entries of $U_L$,
\be
U_L^{i3} = \frac{\lambda_\CN^{i3} \left[ 1 + {\cal O}(\epsilon^2) \right]
}{\left( \left|\lambda_\CN^{13}\right|^2
+ \left|\lambda_\CN^{23}\right|^2 + \left|\lambda_\CN^{33}\right|^2 \right)^{1/2}}
\;\;\;\; , \;\;\;\; i = 1,2~.
\ee
are relevant for atmospheric neutrino oscillations, as discussed
below.
The first two columns of $U_L$ have entries that also depend (to
leading order in $\epsilon$) only on ratios of $\lambda_\CN^{ij}$'s
and on $|\bar{h}_{13}/\bar{h}_{23}|$.
The ratio of the $U_L^{11}$ and $U_L^{12}$ entries, which is relevant for solar
neutrino oscillations, is given by
\be
\left|\frac{U_L^{12}}{U_L^{11}}\right| =
\left|\frac{V_{21} \tilde{l}_3^* - V_{11} \tilde{l}_1 l_{23}}
{V_{11}^* \tilde{l}_3^* - V_{21}^* \tilde{l}_1 l_{23} }\right|
\label{ul}
\ee
This ratio is typically of order unity if most of the
$\lambda_\CN^{ij}$'s have the same order of magnitude.
In particular,
this is true for the phenomenologically favored case discussed below,
which is near the bi-maximal mixing of three neutrinos
\cite{Barger:1998ta}.

%%%%%%%%%%%%%%%%%%%%%%%%%
\subsection{Predictions and experimental constraints}

The neutrino oscillation data constrains the differences of
neutrino squared-masses. With three neutrinos, the atmospheric and
solar neutrino data can be fit nicely (the LSND data cannot be
accomodated).

The atmospheric neutrino data %\cite{Toshito:2001dk}
require a
larger mass splitting compared to the solar data so that
$m_{\nu_3}^2$ is determined to leading order in $\epsilon^2$. The
range for the mass-square difference $(\Delta m^2)_{\rm atm}$ given
by the global fit \cite{Fogli:2001xt} to the data obtained in
atmospheric neutrino experiments and in the CHOOZ reactor
experiment is
\be
m_{\nu_3}^2 \approx (\Delta m^2)_{\rm atm}^{\rm  exp}
= 1.5 - 6.0 \times 10^{-3} \; {\rm eV}^2 \; \; {\rm at \; 99\% \; CL}  ~.
\ee
As long as the combination of Yukawa couplings shown in
Eq.~(\ref{nu3}) is not smaller than unity by many orders of
magnitude, Eq.~(\ref{expo}) requires
\be
\label{exp}
\pi r_c \left(\tilde{M}_{\varphi} - \frac{k}{2}\right) \approx 30 ~,
\ee
where we anticipate that the terms in parenthesis in
Eq.~(\ref{expo}) are of order unity.

The solar neutrino oscillations are controlled by the mass-square
difference of the lighter neutrinos:
\be
(\Delta m^2)_{\rm solar} = m_{\nu_2}^2 - m_{\nu_1}^2 ~.
\ee
Based on the reasonable assumptions that
$|\bar{h}_{23}|, |\bar{h}_{23}| = {\cal O}(1)$,
and that most Yukawa couplings $\lambda_\CN^{ij}$ have
the same order of magnitude, we find the following prediction
for the solar neutrino oscillation scale:
\be
\frac{(\Delta m^2)_{\rm solar}}{(\Delta m^2)_{\rm atm}^{\rm  exp} }
\approx {\cal O}\left( \epsilon^2 \right)~,
\label{atm-solar}
\ee
where $\epsilon$ is given by Eq.~(\ref{epsilon}). In the absence of
fine-tuning, the $\varphi$ scalar mass-squared can be no smaller
than the one-loop contribution in the seven-dimensional theory:
$\delta M_\varphi^2 \sim {\cal O}[\lambda_\varphi M_*^2/(128
\pi^3)]$. Also, we expect that the effective theory description starts breaking down
when $M_\varphi$ approaches $M_*$.
%, we take $M_\varphi^2/M_*^2\lae {\cal O}(10^{-1})$. Therefore,
This leads to an allowed range for $\epsilon$ [see
Eq.~(\ref{epsilon})]:
\be
{\cal O}\left(10^{-3} \sqrt{\lambda_\varphi }\right) \lae \epsilon
\lae {\cal O}\left(10^{-1}/ \sqrt{\lambda_\varphi }\right) ~,
\ee
with the values near the upper end being preferred if the mass
parameter $M_\varphi$ is not too much smaller than the fundamental
scale $M_*$. If $\lambda_{\varphi}$ is of order unity, the ratio
shown in Eq.~(\ref{atm-solar}) is then of order $10^{-2}$.
%We have no reason to expect
%$\lambda_\varphi$ to be much different than order unity, so that
The generic prediction of our model for the solar neutrino scale
then becomes
\be
(\Delta m^2)_{\rm solar}  \approx  10^{-5} - 10^{-4}
\; {\rm eV}^2 ~.
\ee
This prediction fits well the range currently allowed for the
large-mixing angle MSW solution to the solar neutrino problem
\cite{Bahcall:2001cb}.  The large uncertainties due to the unknown
values of various parameters do not allow us to rule out the long
oscillation wavelength solution, whose fit to the data prefers
$(\Delta m^2)_{\rm solar} \approx 10^{-7} \; {\rm eV}^2 $.  On the
other hand, the vacuum oscillation solution requires $(\Delta
m^2)_{\rm solar} \approx 10^{-10} \; {\rm eV}^2$, so that it is
disfavored within our model.

The elements of the unitary matrix $U_L$ are constrained by the
solar, atmospheric and reactor neutrino experiments.
The element $U_L^{13}$ is most tightly constrained by the global
fit to the CHOOZ and solar neutrino data,
which gives $(U_L^{13})^2 < 6.5 \times 10^{-2}$  at 99\% CL
\cite{Bandyopadhyay:2001fb}. Applied to our case, this translates into
a mild restriction:
\be
\frac{\left|\lambda_\CN^{13}\right|}{
\sqrt{\left|\lambda_\CN^{23}\right|^2 + \left|\lambda_\CN^{33}\right|^2 }
} < 0.26  ~.
\label{chooz}
\ee

The atmospheric neutrino data strongly favors pure
$\nu_\mu \leftrightarrow \nu_\tau$
oscillations with a mixing angle satisfying $\sin^2 2\theta_{23} > 0.83$
at the 99\% confidence level \cite{Toshito:2001dk}.
The expression of this mixing angle in terms of the Yukawa couplings
is the same in our model as in the five-dimensional
model of Grossman and Neubert \cite{Grossman:2000ra}, and gives:
\be
0.65 < \left| \frac{\lambda_\CN^{23} }{\lambda_\CN^{33} } \right| < 1.55~,
\ee
This is not a particularly strong constraint, given that the
Yukawa couplings are typically expected to be of the same order of magnitude.
However, if the march of the atmospheric neutrino data towards
{\it maximal} $\nu_\mu \leftrightarrow \nu_\tau$ mixing continues,
then it will be necessary to find a more detailed explanation for why
the ratio of Yukawa couplings shown above is so close to unity.

Finally, the global fits of the large mixing angle
MSW solution to the solar neutrino problem
require \cite{Bahcall:2001cb}
\be
0.2 < \left|\frac{U_L^{12}}{U_L^{11}}\right|^2 \lae 1 ~.
\ee
This constraint is naturally accomodated by order unity values of the
parameters entering the first two columns of the matrix shown in
Eq.~(\ref{ul}).

In summary, we can explain the hierarchy between the electroweak
scale and the scale relevant for the atmospheric neutrino data by
the first exponential factor in Eq.~(\ref{expo}).  We can further explain
the small hierarchy between this scale and the solar neutrino scale
associated with the large-mixing angle MSW solution based on the
typical size of the higher-dimension operators controlling neutrino
flavor mixing.  Furthermore, the currently allowed ranges for the
mixing angles are natural if the various couplings of our model are
of order unity.

%%%%%%%%%%%%%%%%%%%%%%%%%%%%%%%%%%%%%%%%%%%%%%%%%%
\section{Mass Parameters and Warping}

Having observed that the smallness of the neutrino mass scale
relative to the electroweak scale is explained by Eq.~(\ref{expo})
with the exponent given by Eq.~(\ref{exp}), we now comment on the
other mass scales of the model.  These are the scale $M_*$ at which
the seven-dimensional theory, including its six-dimensional
component, breaks down, the parameter $k$ related to the
seven-dimensional cosmological constant, the size $R_u$ of the
universal extra dimensions, and the size $r_c$ of the warped
dimension.  We also introduced the mass $M_{\varphi}$ of the bulk
scalar that localizes the neutrino fields away from the standard
model brane, and the Higgs mass parameter $M_{H}$. These scales
give rise to other, derived scales such as the electroweak scale
$v_h$.

%These scales give rise to other, derived scales which have a
%physical significance, and so it is useful to keep them in mind.
%First and foremost, there is the electroweak scale $v_h$. Second,
%for observers at the standard model brane, the quantity $w(\pi
%r_{c}) M_{*}$ ($=e^{-\pi r_c k} M_{*}$) plays the role of a cutoff,
%since it is the scale at which the graviton KK modes (along the
%warped dimension) become strongly interacting. Associated with this
%quantity is $w(\pi r_{c}) k$, which sets the mass scale for the
%KK modes with respect to the warped dimension.

There are several important relations among the above parameters.
We assume a normalization of the warp factor as in
Eq.~(\ref{form}). (For a different normalization it is only
necessary to interpret $R_u$ in the following formulas as the
proper radius of the universal extra dimensions at the standard
model brane.) The first relation follows from the fact that the
observed standard model gauge couplings, collectively denoted by
$g_{4}$, (as well as the top Yukawa coupling) are of order one.
Writing a typical six-dimensional standard model gauge coupling in
${\cal L}_{\rm SM}$ of Eq.~(\ref{action}) as $g_{6}/M_*$, where
$g_{6}$ is dimensionless, the observed four-dimensional gauge
coupling is given by
\be
\label{4dcouplings}
1 \sim g_{4} = (\pi R_{u} M_* )^{-1} g_{6}~.
\ee
If the standard model gauge interactions become strong at the scale
$M_{*}$, then $g^{2}_{6} \sim 128
\pi^{3}$ \cite{Chacko:1999hg}. In this case, the product $R_{u} M_{*}$
is of order $\sqrt{128\pi}$. For a range of values of $g_6 > 1$,
$M_{*}$ will be somewhat above $1/R_u$, providing a finite range of
validity for the six-dimensional standard model. In the following
we assume that $ {\cal O}(1) < R_{u} M_{*} \lae {\cal O}(10)
$.\footnote{An interpretation based on the AdS/CFT correspondence
\cite{Arkani-Hamed:2000ds} may be useful if $R_{u} M_{*} < 1$, in
which case the present six-dimensional description is not
applicable. We note that the $Z_{8}$ symmetry that lies at the
heart of the remarkable proton decay suppression pointed out in
\cite{Appelquist:2001mj} and which also forbids Majorana masses, is
expected to remain valid. Thus, even though it would not be
possible to talk about a six-dimensional standard model, the
six-dimensional structure still has observable effects in the
low-energy four dimensional theory.}

%It is also possible to imagine that the couplings $g_{6}$ are still
%weak at $M_*$, in which case $\pi R_{u} M_{*}
%\sim g_{6}$ could be of order unity or even much smaller. However,
%if $g_{6} \lae 1$ then $1/R_{u} \gae M_{*}$ and there is no range
%of energies in which the six-dimensional standard model is a valid
%description.\footnote{An interpretation based on the AdS/CFT
%correspondence might be useful in this case
%\cite{Arkani-Hamed:2000ds}.  We note that the $Z_{8}$ symmetry that
%lies at the heart of the remarkable proton decay suppression
%pointed out in \cite{Appelquist:2001mj} and which also forbids
%Majorana masses, is expected to remain valid. Thus, even though it
%would not be possible to talk about a six-dimensional standard
%model, the six-dimensional structure still has observable effects
%in the low-energy four dimensional theory.} In what follows we will
%not consider this possibility and will assume that $ {\cal O}(1) <
%R_{u} M_{*} \lae 20 $.

A second relation follows from naturalness considerations with
respect to the Higgs mass. The Higgs mass parameter $M_{H}$ must be
below $M_*$ for the effective theory description to be valid.
However, on naturality grounds, $M_{H}$ cannot be much smaller than
the one loop corrections. They can be roughly estimated by cutting
off the quartically divergent one-loop integrals, such as the one
arising from the quartic Higgs self-interaction, at the breakdown
scale of the effective, six-dimensional theory. This yields
\be
\label{massparameters}
M_* > M_{H} \gae \delta M_{H} \approx
\sqrt{\frac{\lambda_{6}}{128 \pi^{3}}} M_*~,
\ee
where $\lambda_{6}$ is a dimensionless  coupling in the
six-dimensional theory. Using this estimate, it follows that the
Higgs VEV is given by
\be
\label{VEVineq}
v_{h} = \left[\pi R_{u} M_*
\lambda_{6}^{-1/2}\right] M_{H} \gae
\frac{1}{\sqrt{128 \pi}} (R_{u} M_*)^{2} R_{u}^{-1}~.
\ee
%Since $1/R_{u} \gae 0.5~\mathrm{TeV}$ \cite{Appelquist:2000nn},
%this is perhaps a somewhat tighter upper bound on  $R_u M_*$ than
%above, but still allowing a finite range between $1/R_u$ and
%$M_{*}$.
If $R_u M_* \gae {\cal O}(1)$ then Eq.~(\ref{VEVineq}) can be read
as an upper bound on $1/R_u$ in the TeV range. But the electroweak
precision measurements impose a lower bound $1/R_{u}
\gae 0.5~\mathrm{TeV}$ \cite{Appelquist:2000nn}. It then follows
from $R_{u} M_{*} \lae {\cal O}(10)$ that $M_*$ is in the TeV
range.

Finally we note that the parameters $M_{\varphi}$ and $k$ are
expected to satisfy relations analogous to
Eq.~(\ref{massparameters}) (though, as noted in subsection $5.1$,
we must have $k < M_{\varphi}/3$).  Our discussion of the solar and
atmospheric neutrino data in subsection $6.1$ indicated that
$M_{\varphi}$ must be approximately an order of magnitude below
$M_*$, which is consistent with a naturality estimate analogous to
Eq.~(\ref{massparameters}), if the corresponding coupling
$\lambda_{\varphi}$ is of order unity or smaller.

There are now various possibilities depending on how large the
warping is.  First suppose that the warping is no more than mild:
$e^{\pi r_c k} \sim 1$. In this case, the weakness of gravity must
be attributed to some suppression that lies beyond the
seven-dimensional theory presented here, such as a few other flat
dimensions accessible only to gravity along the lines of
\cite{Arkani-Hamed:1998rs}. It is important to note that the
mechanism for suppressing the neutrino masses presented here is
independent of any such extension. From Eq.~(\ref{exp}), we now
have $\pi r_{c} \tilde{M}_{\varphi} \approx 30$.  With
$\tilde{M}_{\varphi}$ roughly an order of magnitude below $M_*$,
the inverse size of the seventh dimension, $1/\pi r_{c}$, is of
order $10$ GeV.

A more interesting possibility is that $\pi k r_c \gg 1$.  In this
case it is possible to explain the weakness of gravity \`a la
Randall and Sundrum \cite{Randall:1999ee} within the
seven-dimensional model. There are, however, some differences
arising from the existence of the universal extra dimensions.  The
four-dimensional Planck mass is now related to $M_*$ by
\be
\label{PlanckMass}
M_{\rm Pl}^{2} \approx
\frac{M_*^{5} (\pi R_{u})^{2}}{4 k} e^{4 \pi r_c k}~.
\ee
Suppose that $k$ is on the order of (but somewhat less than) $M_*$.
Then if $1/R_{u}$ and $M_*$ are in the TeV range as expected from
the previous considerations,\footnote{Note that we have chosen to
measure all mass scales at the standard model brane with respect to
the corresponding induced metric. Had we measured them with respect
to a metric rescaled by $e^{\pi k r_c}$ as in Ref.
\cite{Randall:1999ee}, we would have concluded that $M_*$ and the
other ``fundamental" parameters of the seven dimensional theory are
of order $M_P$. The cutoff on the effective six dimensional
standard model would, however, remain in the TeV range, being given
by $M_{*} e^{-\pi k r_c}$. None of the physical conclusions
described here would change.} it follows from
Eq.~(\ref{PlanckMass}) that $e^{2
\pi r_c k} \sim 10^{15}$, which translates into $\pi k r_c \sim
20$. (Note that there is an extra factor of two in the exponent
compared to the five-dimensional Randall-Sundrum model.) We then
see from Eq.~(\ref{exp}) that $\pi \tilde{M}_\varphi r_c \sim 40$,
and the inverse size of the seventh dimension, $1/\pi r_{c}$, is
again of order $10$ GeV. The lightest spin-2 KK mode with momentum
along the warped dimension has a mass of about $4 k$, which is
roughly of the same order as the mass of the first KK modes of the
standard model fields. It will be interesting to see which KK modes
will be discovered first if this model is realized in nature.

%With $\pi k r_c \gg 1$, there is a hierarchy of order $w(\pi
%r_{c})^{-1}$ between the ``fundamental" scale $M_*$ and the
%universal KK masses whose scale is set by $1/R_u$ (though no
%hierarchy between $1/M_*$ and the proper size of the universal
%dimensions at the standard model brane location, $w(\pi r_{c})
%R_u$. See Eq.~(\ref{lineelement})). The naturality of such a
%hierarchy can be judged only within the framework of a radius
%stabilization mechanism. For the present purpose, we note only that
%it is not unreasonable to think that the warp factor enters in the
%potential for $R_u$ in such a way as to generate the appropriate
%hierarchy, with all other couplings having their natural values.

%%%%%%%%%%%%%%%%%%%%%%%%%%%%%%%%%%%%%%%%%%%%%%%%%%
\section{Conclusions}

We have presented a higher-dimensional mechanism for generating a
realistic neutrino mass spectrum. The smallness of the neutrino
masses compared with the electroweak scale is explained by an
exponential suppression of the right-handed neutrino wave functions
on the standard-model brane. The hierarchy between the mass scales
associated with the $\nu_\mu \leftrightarrow
\nu_\tau$ and $\nu_e \leftrightarrow \nu_\mu$ transitions,
measured by the atmospheric and solar neutrino data,
respectively, is limited by the effect of flavor-non-diagonal,
higher-dimension operators. As a result, the mass scale of the
solar neutrino oscillations fits well the large-mixing-angle MSW
solution. Furthermore, the neutrino mixing angles are naturally
large if no large hierarchies between the neutrino Yukawa
couplings occur. This is an important result in view of the fact
that a majority of the models in the literature (for a recent
review see \cite{Dorsner:2001sg}) can accomodate only the
small-mixing-angle MSW or vacuum solutions to the solar neutrino
problem, which are less favored by the data. In addition, the
seemingly ``maximal'' mixing required by the atmosperic neutrino
data is consistent with our mechanism for a reasonably large
range of parameters.

While our mechanism has been developed in the framework of the
six-dimensional standard model, it is worth pointing out that it
relies fundamentally on four ingredients that could naturally be
present in a more general class of higher-dimensional theories:\\
$(i)$ Three right-handed neutrinos; \\ $(ii)$ A symmetry structure
forbidding Majorana neutrino masses; \\ $(iii)$ A spatial dimension
compactified on $S^1/Z_2$ and accessible to the right-handed
neutrinos but not to the standard model fields; \\ $(iv)$ A bulk
(effective) scalar field which is odd under the $Z_2$ orbifold
transformation, has a VEV, and couples to the right-handed
neutrinos.

Ingredients $(i)$ and $(ii)$ are automatically present in the
six-dimensional standard model, as required by the six-dimensional
gravitational anomaly cancellation \cite{Dobrescu:2001ae,
Arkani-Hamed:2000hv} and the $Z_8$ rotational symmetry of the two
universal extra dimensions \cite{Appelquist:2001mj}. We were then
led to consider the six-dimensional standard model localized in a
seventh dimension satisfying $(iii)$ and $(iv)$.

More generally, ingredient $(ii)$ could be enforced by lepton
number conservation in a variety of models, and its experimental
test is the absence of neutrino-less double-beta decay.
Ingredients $(iii)$ and $(iv)$ could be present in 4+1
dimensional models, as suggested in Ref. \cite{Georgi:2000wb} as
a possible source of fermion mass hierarchies. We have generalized
this construction by allowing a warped metric.
%The inclusion ($i$) of three right-handed neutrinos is
% also possible in 4+1 dimensions.
We have analyzed the conjectured restriction on
the number of gauge-singlet fermions \cite{Grossman:2000ra}, and
have found that
% in 4+1 dimensions there are
% there are no gravitational anomalies and therefore
in 6+1 dimensions
the local and global gravitational anomalies
cancel independently of the number of right-handed neutrinos,
provided the fermion content is free of six-dimensional anomalies.
(In 4+1 dimensions no restrictions arise because there are
no {\it local} or {\it global} gravitational anomalies in four or five dimensions.)

Finally, it is worth noting that the model presented here has
another intriguing feature. In the most appealing version of the
model, the extra dimension that leads to an exponential
suppression of the right-handed neutrino wave functions also
solves the hierarchy problem along the lines of Randall and
Sundrum \cite{Randall:1999ee}, while explaining proton stability
based on six-dimensional Lorentz invariance as in
Ref.~\cite{Appelquist:2001mj}. In this case collider searches at
the TeV scale will reveal graviton Kaluza-Klein modes in addition
to the Kaluza-Klein modes of the standard model fields.

%%%%%%%%%%%%%%%%%%%%%%%%%%%%%%%%%%%%%%%%%%%%%%%%%%%%%%%%%%%%%%%%%
%%%%%%%%%%%%%%%%%%%%%%%%%%%%%%%%%%%%%%%%%%%%%%%%%%%%%%%%%%%%%%%%%
\bigskip\bigskip\medskip

%\noindent
{\bf Acknowledgements:} \ We thank H.-C.~Cheng, S.~Glashow,
M.~Luty, I.~Mocioiu, R.~Mohapatra, and H.~Murayama for useful
conversations and communications. We acknowledge the hospitality of
the Aspen Center for Physics during early stages of this work. This
work was supported by DOE under contract DE-FG02-92ER-40704.

%%%%%%%%%%%%%%%%%%%%%%%%%%%%%%%%%%%%%%%%%%%%%%%%%%
\section*{Appendix}
\renewcommand{\theequation}{A.\arabic{equation}}
\setcounter{equation}{0}

In this Appendix we prove that the real scalar $\varphi$ has a nonzero
VEV whenever the parameters $k$, $M_\varphi$ and $r_c$ satisfy the
conditions (\ref{bound}) and (\ref{bound2}). The
rescaled VEV $\f(z) =
\sqrt{\lambda_\varphi} \, M_\varphi^{-1} f(z) $
must satisfy the $z$--independent equation
\be
\label{VEV3}
\frac{d^2 \f}{d z^2}
= 6k \frac{d \f}{d z} + M_\varphi^2 \left( \f^3 - \f \right)~.
\ee
This equation describes the mechanical motion in the potential
$-  M_\varphi^2 (\f^2 - 1)^{2}/4$, and in
the presence of an \textit{anti}--friction term proportional to $k$.
The $z$ coordinate plays the role of ``time" variable.

We will prove the existence of a solution to (\ref{VEV3}), satisfying
the boundary conditions $\f(0)=\f(\pi r_c)=0$, by analyzing the ``flows"
in the equivalent first order system:
\bear
\label{flow}
\frac{d \f}{d z}&=&P_{\f} ~, \nonumber \\ [.5em]
\frac{d P_{\f}}{d z}&=&6k P_{\f}+M_{\varphi}^2\left(\f^3-\f\right)~.
\eear
In the $(\f,P_{\f})$ plane, the solution sought
corresponds to a flow that starts somewhere on the $\f=0$ line and,
after time $\pi r_c$, comes back to this line (because of the
symmetry $\f \rightarrow -\f$, we can restrict $\f \ge 0$.)
More precisely, we shall prove that for any $k
< M_{\varphi}/3$, there are flows starting and ending on the $\f=0$
line, with total elapsed times ($\pi r_c$) ranging from
$\pi / \omega$ to infinity, where
\be
\omega \equiv \sqrt{ M_{\varphi}^2 - 9k^2 } ~.
\ee

The system (\ref{flow}) has two fixed points,
$(\f,P_{\f})=(0,0)$ and $(1,0)$, near which it
may be linearized.
In the vicinity of $(0,0)$, Eqs.~(\ref{flow})
has the following explicit solution
satisfying the initial conditions $(\f,P_{\f})=(0,P_0 >0)$:
\bear
\f(z)&=&\frac{P_0}{\omega}\:e^{3kz}\: \sin\omega z ~,
\nonumber \\ [.5em]
P_{\f}(z)&=&
\f(z) \left( 3k  + \omega \tan^{-1} \omega z \right) ~.
\label{sol1}
\eear
Since $\f$ vanishes at $z=\pi/\omega$, the existence of solutions
is established for a  separation between the branes of $\pi r_c
\rightarrow \pi/\omega$ [see Eq.~(\ref{bound2})], at least when
$P_0 \rightarrow 0$ such that the linear approximation is reliable.

\begin{figure}[t]
\begin{center}
%\scalebox{1.2}[1]{
\hspace*{-1.3cm}\includegraphics{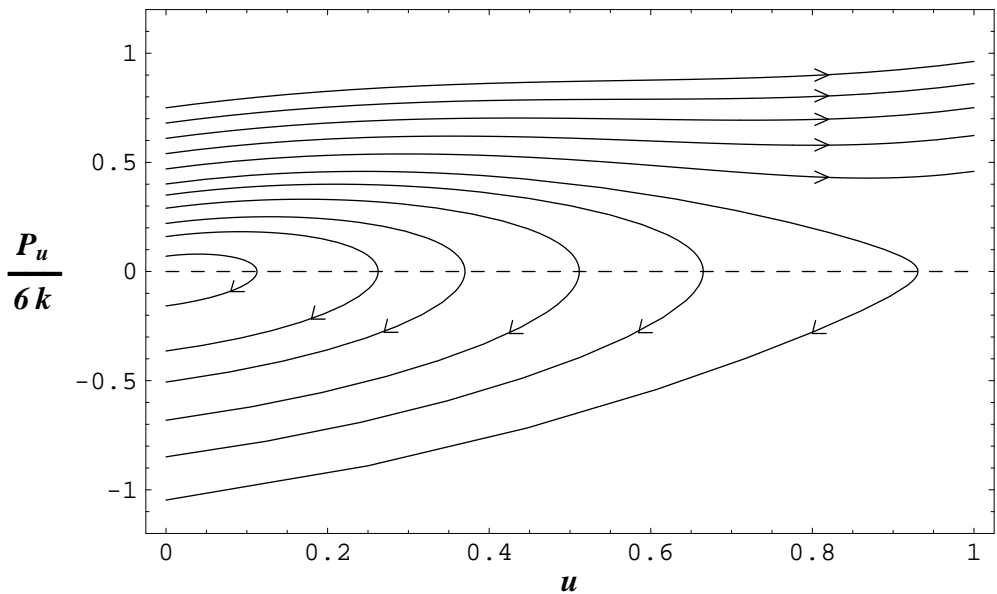} %}
\par
\vskip-2.0cm{}
%\includegraphics*[5cm,11.52cm][18.0cm,19.8cm]{flow.ps}
%\par
%\vspace{-.8cm}
%\vskip-1.16cm{}
\end{center}
\caption{\small The flow diagram when $M_{\varphi}/3k=4$.}
\label{appfig}
\end{figure}

A similar linear analysis around $(1,0)$ shows that there is an
attractive flow that approaches the fixed point $(1,0)$ from the
region $\f<1$, $P_{\f}>0$, and a repulsive flow in the region $\f<1$,
$P_{\f}<0$.
For either one of them, the total time is infinite.
This shows that adjacent flows that come arbitrarily close to
$(1,0)$ will spend an arbitrarily long time in its vicinity.
Note that
whenever the flow enters the region $\f<1$, $P_{\f}<0$, it will be
driven to the $\f=0$ line, since the only possibility is to roll down
the slope of the potential.

We now prove the possibility of reaching these flows from a point
$(0,P_{\f}>0)$, based on the following remarks. First,
Eq.~(\ref{flow}) shows that along the line $P_{\f}=0$, and for
$\f<1$, the trajectories flow vertically downward (see
Fig.\ref{appfig}).  Second, the trajectories passing through the
$\f=1$ line flow into the region $\f>1$ whenever $P_{\f}>0$, and
then they are driven to $\f \rightarrow \infty$ due to the negative
slope of the potential. Tracing these trajectories back in time, it
follows from the first remark that they necessarilly cross the
$\f=0$, $P_{\f} \geq 0$ half-line. However, they cannot cross the
origin because the flows cannot stop except at fixed points, and
the point (0,0) cannot be reached [for small enough $P_{\f}>0$ the
trajectories stay only in the vicinity of $(0,0)$, see
Eqs.~(\ref{sol1})]. Note that the flows cannot cross due to the
uniqueness of the solutions to ordinary differential equations such
as Eq.~(\ref{flow}). Third, tracing back in time the trajectory
attracted to the fixed point $(1,0)$, the same argument as above
shows that it crosses the line $\f=0$, $P_{\f}>0$. This
``critical'' trajectory, which starts at some critical $(0,P_c>0)$
and is attracted to $(1,0)$, sets a boundary between qualitatively
different flows. Finally, from any point $(\f,0)$, $\f<1$, the
trajectory traced back in time crosses the line $\f=0$ at some
$0<P_{\f}<P_c$. Thus, a point $(\f,0)$ with $\f$ arbitrarily close
to $1$ corresponds to a trajectory that spends an arbitrarily long
time in the vicinity of $(1,0)$.

We have shown so far that there are solutions for $\pi r_c$ close
to either $\pi/\omega$ or infinity. By continuity, it follows that
there are solutions to Eq.~(\ref{VEV3}), satisfying the boundary
conditions $\f(0)=\f(\pi r_c)=0$, for any $r_c > 1/\omega$. This
completes the proof that Eqs.~(\ref{bound}) and (\ref{bound2}) are
necessary and sufficient conditions for the existence of a nonzero
$\varphi$ VEV.

%%%%%%%%%%%%%%%%%%%%%%%%%%%%%%%%%%%%%%%%%%%%%%%%%%%%%%%%%%%%%%%%%%%
 \vfil \end{document}